\documentclass[aps,pra, twocolumn,superscriptaddress]{revtex4-2}
\usepackage{graphicx} 
\usepackage{xcolor}

\DeclareUnicodeCharacter{2212}{-}
\usepackage{physics}
\usepackage{leftindex}
\usepackage{upgreek}
\usepackage{amsmath}
\usepackage{amsfonts}
\usepackage{bbm}
\usepackage[normalem]{ulem}

\newcommand{\corr}[1]{#1}
\newcommand{\ifunam}{Instituto de F\'isica, Universidad Nacional Aut\'{o}noma de M\'{e}xico, Ciudad de M\'{e}xico 04510, M\'exico}

\newcommand{\uama}{\'Area de F\'isica Te\'orica y
Materia Condensada, Universidad Aut\'onoma Metropolitana
Azcapotzalco, Ciudad de M\'exico 02200, M\'exico}
\newcommand{\iimas}{Instituto de Investigaciones en Matem\'aticas Aplicadas y en Sistemas, Universidad Nacional Aut\'{o}noma de M\'{e}xico, Ciudad de M\'{e}xico 04510, M\'exico}

\begin{document}
\title{Collective coupling of driven multilevel atoms and its effect on four-wave mixing}

\author{P.~Yanes-Thomas}
\affiliation{ \ifunam}
\author{R.~Guti\'{e}rrez-J\'auregui}
\affiliation{ \ifunam}
\author{P.~Barberis-Blostein}
\affiliation{ \iimas}
\author{D.~Sahag\'un S\'anchez}
\affiliation{ \ifunam}
\author{R.~J\'auregui}
\email[Email:]{rocio@fisica.unam.mx}
\affiliation{ \ifunam}
\author{A.~Kunold}
\email[Email:]{akb@azc.uam.mx}
\affiliation{ \uama}

\begin{abstract}
Microscopic models based on multilevel atoms are central to optimizing non-linear optical responses and the coherent control of light. These models are traditionally based on single-atom effects that are parametrically extrapolated to include collective effects, such as an enhanced response or propagation within atomic media. In this work we present a systematic analysis of the cooperative effects arising in driven systems composed of multilevel atoms coupled via a common electromagnetic environment. The analysis is based on an interplay between dressed states induced by the driving field and photon exchanges, and collective decay channels. This theory is applied to the case of four-wave mixing induced by a pair of lasers acting on {an atomic pair} with internal levels in the diamond configuration. The effect of inter-atomic correlations and collective decay over the photons created in this nonlinear process is then explored.
  We identify three regions of operation: (i) laser-dominated; (ii) intermediate; and (iii) dipole-dominated. The dependence of single and two-photon correlations are studied in detail for each region by varying atomic orientations and laser parameters { consistent with current experiments involving atomic gases.}Photonic correlation functions are shown to exhibit a transition from a Lorentz-like dependence on the two-photon detuning---with general features that can be obtained in an isolated atom scheme---to a two-peaked distribution when the dipole-dipole interactions become relevant. For weak Rabi frequencies whose value is smaller than the highest collective decay rate, the atoms are trapped inside their ground state as they approach each other. It is found that the anisotropy of the dipole-dipole interaction and its wave nature are essential to understand the behavior of the photons correlations. Signatures of these processes are identified for existing experimental realizations. The intuition obtained from this connection helps to uncover relevant parameters that could be exploited for quantum control protocols based on dispersive and dissipative cooperative effects in multilevel systems.
\end{abstract}

\maketitle

\section{Introduction}

The optical response of an atomic medium is modified by external electromagnetic fields. Intense laser fields, for example, have been shown to alter this response directly by creating excitation pathways between electronic states or indirectly by manipulating the atomic positions. Interference among these excitation pathways alters the interaction between atoms and their electromagnetic environment, resulting in dressed states of light and matter that can be controlled in real time to create transparency windows in otherwise opaque media~\cite{Boller_1991,Juzeliunas_2002,Fleischhauer_2005,Finkelstein_2023}, promote non-linear responses~\cite{Alexanian_2006, Ardelt_2016, Bounuar_2017} or, for molecules, tune their interaction~\cite{Yan_2020}. Control over  atomic positions, by comparison, can be used to promote coherent scattering between the atoms. Atoms inside dense ensembles display a collective response that is manifested as altered scattering rates~\cite{Gross_1982} and directional scattering~\cite{Bromley_2016}, which are accompanied by modified resonant lines~\cite{James_1993,Keaveney_2012,Javanainen_2014,Jenkins_2016}.  The effects become more pronounced for ordered arrays~\cite{Bettles_2016,Rui_2020,Javanainen_2020,Masson_2022}. It is this ability to fully control different atomic degrees-of-freedom with high precision that has turned atomic systems into fundamental tools for creating versatile light matter interfaces.

As experiments scale in both size and complexity, an increasingly rich interplay between the control of external and internal degrees-of-freedom is observed. In a regime characterized by strong fields and dense atomic media, atomic states are dressed by both external { and internal} fields { including} those scattered by neighboring atoms to create a collective response. Coherent aspects remain~\cite{Jenkins_2013}, allowing the tailoring of exotic responses as non-reciprocal~\cite{GJ_2022} or topological materials~\cite{Cooper_2019,Leseleuc_2019, Semeghini_2021,Ohler_2023},  with natural extensions to ensembles of artificial atoms and emitters~\cite{Hadad_2010,Hadad_2013}. The control can also be used to explore the connection between macroscopic properties of a material with microscopic models, thus allowing the characterization of the properties of the coherent scattering into compact, macroscopic parameters such as the optical density~\cite{Eberly_1969}. Such an approach extends to modern technologies, which place a premium on identifying the relevant variables for controlling the optical response in a nonlinear setting.

We present here a systematic approach for addressing the interplay between coherent and dissipative collective effects in atomic media, and apply it to the particular scenario of four-wave mixing in a cold atomic cloud where two-body interactions dominate. The approach begins from an ensemble of multi-level atoms immersed in an electromagnetic background that can include external fields.

The dynamical equations describing the atomic and electromagnetic field operators are obtained as a driven and multi-level generalization of~Ref.~\cite{Lehmberg_1970}, that can readily incorporate non-linear optical processes. As a specific example, we consider the case of two four-level atoms driven by pump lasers. This case provides the basic ingredients needed to describe non-linear and collective effects in an atomic ensemble---either ordered or disordered---where pair-wise interactions become prominent. The response of the system is seen to change as we transition between laser-dominated and dipole-dominated regimes. The transition is captured by the structure of the collective dressed states of the system, which are then used to characterize the photons scattered in this non-linear process. We show that the collective effects give way to self-seeding, a line-splitting in the atomic transitions,and a change of the emission rate of the outgoing photons.
As a consequence, we identify conditions where the detailed structure of the electromagnetic response can not be obtained using standard macroscopic models. Our approach is directed towards identifying the relevant variables for controlling the collective response of an atomic ensemble via external electromagnetic fields with potential applications in quantum control and engineering.

The paper is organized as follows. In Section \ref{sec:model} we introduce a microscopic model for an atomic ensemble driven by external laser fields. We obtain dynamical equations for the emitters operators, placing emphasis on the effective interactions induced by the laser fields over multi-level atomic dipoles and the effect of dissipation. Then, in Sec.~\ref{sec:twofourlevatom}, we treat the explicit example of two four-level atoms using  experimental parameters for cold atomic clouds. We obtain the dressed-state of the system and organize them into separate blocks caused by particular selection rules. Three interaction regimes are identified. An analysis of how this collective response is imprinted on the outgoing photons in the three regimes is presented in Sec.~\ref{subsec:g2}. Using the collective dressed states, we identify the relevant variables that determine the properties of the outgoing photons. Section~\ref{sec:Conclusions} is left for summary and conclusions, where we highlight signatures of collective processes in four-wave mixing that can be found within current experiments.

\section{Multi-level emitters coupled to a common environment}\label{sec:model}

Consider an ensemble of $n_a$ identical, non-overlapping emitters that are coupled to a common electromagnetic environment. Each emitter is characterized by its position $\boldsymbol{r}_{\alpha}$ and is composed of $n_{l}$ internal levels $\ket{m}_{\alpha}$ of corresponding energies $\varepsilon_{m}$ with $\alpha = \lbrace 1,2,\dots, n_a \rbrace$ and $m= \lbrace 0,1, \dots, n_{l}-1\rbrace$. The environment is described by the electric field operator $\mathbf{E}$, which reads
\begin{equation} \label{eq:field_op}
\mathbf{E}(\mathbf{r},t) =\sum_{q}\mathcal{E}_{q} \left[\boldsymbol{e}_q
 \mathrm{e}^{i\boldsymbol{k}_{q}\cdot\boldsymbol{r}}a_{q}(t)
  +\boldsymbol{e}^*_q\mathrm{e}^{-i\boldsymbol{k}_{q}\cdot\boldsymbol{r}}a_{q}^\dagger (t)
  \right] \, ,
\end{equation}
when decomposed into a plane-wave expansion. Each plane wave is characterized by a wave vector $\boldsymbol{k}_q$, angular frequency $\omega_q =c\sqrt{\boldsymbol{k}_q\cdot\boldsymbol{k}_q }$,  and polarization unit vector $\boldsymbol{e}_q$. These traits are encompassed in an index $q$ for simplicity. The mode energies are normalized by the vacuum field amplitude $\mathcal{E}_{q}=\sqrt{ \hbar \omega_q/2\epsilon_0 \mathcal{V}}$, with $\mathcal{V}$ the spatial volume confining the field, while creation $a_{q}^\dagger$ and annihilation operators $a_{q}$ obey the commutation relation $[a_{q},a_{q^\prime}^\dagger]= \delta_{q,q^\prime}$.

The emitters are assumed to move slowly, such that no appreciable changes occur in their position in the time required to modify their collective internal states. We then remove this motion from our analysis and describe this light-matter system by the Hamiltonian
\begin{equation}
\mathcal{H}= \mathcal{H}_A + \mathcal{H}_P + \mathcal{H}_{AP} \, , \label{eq:hamiltonian}
\end{equation}
where
\begin{eqnarray}
\mathcal{H}_A &=& \sum_{\alpha=1}^{n_a} \sum_{m=0}^{n_l-1}  \varepsilon_{m}\sigma_{m,m}^{\alpha} \, , \label{eq:free_at}\\
\mathcal{H}_P &=& \sum_q\hbar \omega_qa_q^\dagger a_q \, , \label{eq:free_ph}
\end{eqnarray}
are free terms for emitters and field, respectively. They are coupled by the dipolar interaction
\begin{equation}
    \mathcal{H}_{AP} =-\sum_{\alpha =1}^{n_{a}} \sum_{n,m=0}^{n_{l}-1}\mathbf{E}(\mathbf{r}_{\alpha})\cdot \boldsymbol{s}_{mn}^{\alpha} \, ,
\end{equation}
allowing for each emitter to probe its local electric field through its dipole moment operator $\mathbf{p}$. The interaction includes both rotating and counter-rotating terms via
\begin{equation}
\boldsymbol{s}_{mn}^{\alpha}(t)=\boldsymbol{p}_{mn}\sigma_{mn}^{\alpha}
+\boldsymbol{p}_{mn}^*\sigma_{mn}^{\alpha\dagger}.
\end{equation}
with $\boldsymbol{p}_{mn}=\left\langle m \left\vert  \mathbf{p}\right\vert n \right\rangle$ the matrix element connecting two internal states and $\sigma_{mn}^{\alpha}=\sigma_{nm}^{\alpha \dagger}=\ket{m}_{\alpha} \leftindex_{\alpha}{\bra{n}}$ atomic operators that signal transitions within the internal levels of a single emitter and thus commute for any two distinct emitters, $[\sigma_{mn}^{\alpha}, \sigma_{m^\prime n^\prime}^{\beta}] = 0$ for $(\alpha \ne \beta)$.

\subsection{Dynamical equations}\label{subsec:dynamical_equations}

The field mediates the interaction between the emitters and introduces boundary conditions via external drives and collective decay channels. We now include these effects over arbitrary emitter operators placing emphasis on the distinctions caused by the multi-level structure of the emitters.

We begin with the Heisenberg equations for the field operators $a_{q}(t)$. Under the action of $\mathcal{H}$, these operators evolve as
\begin{equation}
\dot{a}_{q}= -i\omega_q a_{q} +\frac{i}{\hbar}   \mathcal{E}_{q} \sum_{\alpha}\sum_{n,m}
\boldsymbol{e}^*_{q}\cdot \boldsymbol{s}_{mn}^{\alpha}
\mathrm{e}^{-i\boldsymbol{k}_{q}\cdot\boldsymbol{r}_\alpha},\label{eq:aqdifeq}
\end{equation}
with formal solution
\begin{align}
a_{q}(t&)
=a_{q}(0)\mathrm{e}^{-i\omega_q t} \nonumber \\
&+\frac{i}{\hbar}\mathcal{E}_{q}\sum_{\alpha,n,m}
\int_0^t dt^\prime \mathrm{e}^{i\omega_q (t^\prime-t)-i\boldsymbol{k}_{q}\cdot\boldsymbol{r}_\alpha}
\boldsymbol{e}^*_{q}\cdot \boldsymbol{s}_{mn}^{\alpha}(t^\prime).
\label{eq:aq1}
\end{align}

When substituted into Eq.~(\ref{eq:field_op}), this solution naturally divides the electric field into free and scattered terms $\mathbf{E} = \mathbf{E}_{\text{f}} +  \mathbf{E}_{\text{s}}$. Their positive frequency components are
\begin{flalign}
\mathbf{E}^{(+)}_{\text{f}}(\mathbf{r},t) = \sum_q
\mathcal{E}_{q} \boldsymbol{e}_{q} \mathrm{e}^{i\left(\boldsymbol{k}_{q}\cdot\boldsymbol{r} -\omega_q t \right)}a_q(0) \, , \label{eq:Free_fields} \\
\mathbf{E}^{(+)}_{\text{s}}(\mathbf{r},t) = \frac{i}{\hbar} \sum_q
\mathcal{E}_{q}^{2}\sum_{\alpha,n,m} \mathrm{e}^{i\boldsymbol{k}_{q}\cdot(\boldsymbol{r} - \boldsymbol{r}_\alpha)}  \boldsymbol{e}_{q} \, \nonumber \\
\times\int_0^t dt^\prime \mathrm{e}^{i\omega_q (t^\prime-t)} \boldsymbol{e}^*_{q}\cdot \boldsymbol{s}_{mn}^{\alpha}(t^\prime). \label{eq:scattered_fields}
\end{flalign}

The scattered term describes the field radiated by the emitters. In free-space, where the emitters couple to a large number of modes, the sum over $q$ turns into an integral over all possible frequencies $d\omega$ and propagation angles $d\Omega_{\mathbf{k}}$, and a sum over polarization vectors $\boldsymbol{e}_{q;\mathbf{k}}$. For weak enough coupling between field and emitters, the coupling strength and mode densities can be evaluated at the resonance frequency 
\begin{equation}
\Delta_{mn}=(\varepsilon_n-\varepsilon_m)/\hbar \, ,
\end{equation}
for each transition. Under this first Markov approximation~\cite{Gardiner_1985}, the integrals are readily performed to give

\begin{align}\label{eq:scatt_Markov}
    \mathbf{E}^{(+)}_{s}(\mathbf{r}_{\beta},t) =  & \sum_{\alpha} \sum_{\lbrace n,m \rbrace} \left(  \frac{\Delta_{mn}^{2}}{\epsilon_{0}c^{2}} \right) \overleftrightarrow{\mathbf{G}}(\mathbf{r}_{\alpha\beta},\Delta_{mn})\cdot \mathbf{p}_{n,m} \nonumber \\
    \times & \mathrm{e}^{-i\Delta_{nm}{t}_{\alpha\beta}} \sigma_{mn}^{\alpha}(t-t_{\alpha\beta}) \Uptheta(t-t_{\alpha\beta}) \, ,
\end{align}
where the sum $\lbrace n,m \rbrace$ is taken over positive frequencies $\Delta_{mn}$ only and $ct_{\alpha \beta} = \vert \mathbf{r}_{\alpha\beta}\vert$ describes the time delay between the emission of a photon at $\mathbf{r}_{\alpha}$ and its detection at $\mathbf{r}_{\beta}$,  with $\mathbf{r}_{\alpha\beta}=\mathbf{r}_{\alpha}-\mathbf{r}_{\beta}$. The field propagation is then dictated by the Heaviside function $\Uptheta$ and Green function

\begin{equation}
 \overleftrightarrow{\mathbf{G}}(\mathbf{r},\omega) = \frac{\omega}{4 \pi c}\frac{e^{i \xi}}{\xi^{3}}\left[(\xi^{2} + i \xi -1) \mathbbm{1} -(\xi^{2} + 3i \xi -3) \frac{\mathbf{r}\otimes \mathbf{r}}{r^{2}} \right],
\label{eq:tensorF}\end{equation}
with $\xi = \omega \vert \mathbf{r}\vert /c$ a natural dimensionless distance. 

By evaluating emitter and field operators at different times, Eq.~(\ref{eq:scatt_Markov}) describes the evolution of a light source as its output propagates towards a target. The spatial separation between source and target is effectively removed by transforming into an interaction picture where the sources are retarded. For a small ensemble, where the only changes in the emitter as the photon propagates from one end of the ensemble to the other are given by its free evolution, this retardation results in a local phase~\cite{Lehmberg_1970}. The scattered field under this assumption is

\begin{align}
    \mathbf{E}^{(+)}_{s}(\mathbf{r}_{\beta},t) = \sum_{\alpha} \sum_{\lbrace n,m \rbrace} & \left(  \frac{\Delta_{mn}^{2}}{\epsilon_{0}c^{2}} \right) \overleftrightarrow{\mathbf{G}}(\mathbf{r}_{\alpha\beta},\Delta_{mn})\cdot \mathbf{p}_{nm} \nonumber \\
    \times &\sigma_{mn}^{\alpha}(t) \Uptheta(t) \, \label{eq:positive},
\end{align}

We now turn our attention to arbitrary operators $Q(t)$ of the emitters. Under the action of $\mathcal{H}$ and within the rotating wave approximation these operators evolve as
\begin{flalign}\label{eq:Qdot2}
\dot{Q}=
\frac{i}{\hbar}
    \left[\mathcal{H}_{A},Q\right] - \frac{i}{\hbar} &\sum_{\alpha}\sum_{\lbrace n,m \rbrace} \left[ \sigma_{mn}^{\alpha},Q \right] \left(\mathbf{p}_{mn} \cdot \mathbf{E}^{(+)}(r_{\alpha})\right) \nonumber \\
    - &\left(\mathbf{p}_{mn}^{*} \cdot \mathbf{E}^{(-)}(r_{\alpha}) \right) \left[Q, \sigma_{mn}^{\alpha \dagger} \right] ,
\end{flalign}
where all operators are evaluated at time $t$. For simplicity, we consider that the free internal energies $\epsilon_{m}$ are already Lamb shifted~\cite{Agarwal_1973}. 

By inserting the fields from Eqs.~(\ref{eq:Free_fields}-\ref{eq:scattered_fields}) into Eq.~(\ref{eq:Qdot2}), the dynamical equations are obtained following the standard approach~\cite{Gross_1982,Gardiner_1985}. Here, terms that involve operators $\sigma_{mn}^{\alpha}$ and $\sigma_{m^\prime n^\prime}^{\beta}$ are removed unless the separation between their transition frequencies, $\vert \vert \Delta_{mn}\vert - \vert \Delta_{m^\prime n^\prime}\vert \vert$, is smaller than the natural linewidths of the relevant internal levels. After removing these fast-oscillating terms and performing the summations, Eq.~(\ref{eq:Qdot2}) becomes
\begin{equation}
\dot{Q}= -(i\hbar)^{-1}\left[\mathcal{H}_{A}+ \mathcal{H}_{E} + \mathcal{H}_{dd},Q\right]  +\mathcal{L}\left[Q(t)\right], \label{eq:Qdot13}
\end{equation}
where the effects of the free fields and dipole-dipole interactions are accounted for through
\begin{flalign}
\mathcal{H}_{E} &= -\sum_{\alpha} \sum_{\lbrace n,m \rbrace} {\sigma}_{mn}^{\alpha} \mathbf{p}_{mn} \cdot \mathbf{E}^{(+)}_{\text{f}}(\mathbf{r}_{\alpha},t) + \text{H.c.}  \, , \label{eq:free_fields_Ham}\\
\mathcal{H}_{dd} &=-\sum_{\alpha \ne \beta} \sum_{\lbrace n,m \rbrace}{\sum_{\lbrace n^\prime,m^\prime \rbrace}}\hbar\Omega^{\alpha \beta}_{m,n;m^\prime, n^\prime} \sigma_{mn}^{\alpha\dagger} \sigma_{m^\prime n^\prime}^{\beta}\, . \label{eq:Rabi_intro}
\end{flalign} 
Emission of photons by atoms is included via the Lindblad term
\begin{equation}
\mathcal{L}\left[\bullet\right]= \tfrac12\sum_{\alpha \beta} \sum_{\lbrace n,m \rbrace}{\sum_{\lbrace n^\prime,m^\prime \rbrace}}
  \gamma^{\alpha \beta}_{m,n;m^\prime, n^\prime} \mathfrak{L}_{mn;m^\prime n^\prime}^{\alpha \beta}\left[\bullet\right],
  \label{eq:similar_energies}
\end{equation}
with $
\mathfrak{L}_{mn;m^\prime n^\prime}^{\alpha \beta}\left[\bullet\right]=2\sigma_{mn}^{\alpha\dagger} \bullet \sigma_{m^\prime, n^\prime}^{\beta} -\left\{\sigma_{mn}^{\alpha\dagger}
\sigma_{m^\prime n^\prime}^{\beta} , \bullet \right\} .$

The dipole-dipole coupling $\Omega^{\alpha \beta}$ and the decay rates  $\gamma^{\alpha \beta}$ are related to the imaginary and real parts of the Green function by

\begin{align}
\Omega^{\alpha \beta}_{m,n;m^\prime, n^\prime} + &i\tfrac12 \gamma^{\alpha, \beta}_{m,n;m^\prime, n^\prime} = \left(\frac{3 \pi c \Gamma_{mn;m^\prime n^\prime} }{\Delta_{m^{\prime}n^{\prime}}}\right)  \nonumber \\
& \times \hat{\mathbf{p}}_{mn}^{*}\cdot\overleftrightarrow{\mathbf{G}}(\mathbf{r}_{\alpha\beta},\Delta_{m'n'}) \cdot \hat{\mathbf{p}}_{m^\prime n^\prime} \, , \label{eq:Green_Function}
\end{align}
where $\hat{\mathbf{p}}_{m^\prime n^\prime}$ and $\hat{\mathbf{p}}_{mn}^{*}$ are unitary vectors pointing  along the direction of the dipole moments and
\begin{equation}
    \Gamma_{mn;m^\prime n^\prime} = \left(\frac{1}{4\pi \epsilon_{0}} \right) 
    \frac{4\Delta^{3}_{mn} \left\vert \mathbf{p}_{mn}^*\right\vert
    \left\vert\mathbf{p}_{m^\prime n^\prime}\right\vert}{3\hbar c^{3}} \, .
\end{equation}
When the electric dipole moments coincide, this rate corresponds to the spontaneous decay rate between levels $m$ and $n$ of an individual atom $$\Gamma_{mn;m n}=: \Gamma_{mn} \, . $$

\subsection{Dressed states and collective decays}\label{sec:dressedstates}

Equation~(\ref{eq:Qdot13}) describes the evolution of an arbitrary operator of the emitter ensemble. It is composed of three competing terms that determine the ensemble response: (i) free modes carrying the external fields; (ii) coherent dipole-dipole coupling; and (iii) decay including collective effects. 

The free modes and dipole-dipole coupling compose the Hamiltonian evolution. For coherent sources such as a laser, the relevant free modes $a_{q}$  in Eq.~(\ref{eq:free_fields_Ham}) carry the laser photons and are thus replaced by state amplitudes $\alpha_{q}$ to give~\cite{Carmichael_1999} 
\begin{equation}
\mathcal{H}_{E}(t)=-\hbar\sum_\alpha\sum_{mn}\sigma_{mn}^\alpha\Lambda_{mn} \mathrm{e}^{i\boldsymbol{k}\cdot\boldsymbol{r}_\alpha} \mathrm{e}^{-i\omega_{\ell} t} +\text{H.c.}
\label{eq:hr2}
\end{equation}
where $\omega_{\ell}$ is the laser frequency, and $\Lambda_{mn} = \alpha_{\ell}\mathcal{E}_{\ell} \mathbf{e}_{\ell}\cdot\mathbf{p}_{mn}/\hbar$ represents the Rabi frequency. The driving fields create coherent pathways that connect two internal emitter states whose transition frequencies are near the laser frequency $(\omega_{\ell}\simeq\Delta_{mn})$. This dressing of the emitter states has been famously used to introduce transparency windows in otherwise opaque media~\cite{Boller_1991,Juzeliunas_2002,Fleischhauer_2005,Finkelstein_2023}. The dipole-dipole coupling, by comparison, allows for the exchange of excitations between two emitters. The exchange rate depends on the emitter separation via $\Omega^{\alpha, \beta}$ and is responsible of imposing relative phases between the emitters and shifting their energy levels. The complete Hamiltonian including drive and dipole-dipole coupling can be diagonalized to obtain the collective dressed states of the ensemble. The structure of these states is ultimately influenced by the competition of both interactions and changes radically with the emitter separation. 

The final influence on the ensemble comes from atomic radiative decay,
which introduces damping into the system through collective channels. The channels are formed from the interference of the individual decay paths obtained by diagonalizing the $\gamma$ matrix of Eq.~(\ref{eq:similar_energies}). Destructive interference gives way to sub-radiant channels whose eigenvalue is smaller than the corresponding individual decay rate, while constructive interference leads to super-radiant ones whose eigenvalues are larger than the individual decay rate. For multi-level emitters collective and individual channels can coexist, as decay paths associated to large transition wavelengths may interfere at distances where those of shorter wavelengths do not~\cite{Masson_2024}. In the same sense, for transitions sharing similar wavelengths spontaneous emission can lead to coherence between different states~\cite{Javanainen_1992,Anton_2005,Zhu_2016}. 

The phases imposed on the emitters inside the collective dressed states are usually different than those imposed during the collective decay, as the matrices $\gamma$ and $\Omega$ do not commute with one another outside of restricted cases~\cite{Gross_1982}. Traditional approaches { for modeling} the optical response of the ensemble tend to work inside regimes where a particular behavior dominates~\cite{Olmos_2013,Zhu_2016},  { and} describe its response in the relevant basis  including the remaining terms numerically or as a perturbation. In the following sections we consider a particular example where this interplay between coherence and decay is parametrically changed by an external field and emitter separation. For { close  emitters}, it allows us to delimit three regions that can be accessed by varying the external laser fields opening the possibility of further characterizing properties such as density or optical depth of { an} ensemble.

\section{Two four level atoms}\label{sec:twofourlevatom}
The formalism described above can be applied both to ordered arrays or disordered atomic ensembles. We now focus on the particular case of two multi-level atoms driven by coherent fields in a relevant situation for nonlinear optics. By focusing on the two atom scenario we are able to characterize the essential features introduced by the collective effects that can alter the nonlinear response. This scenario is relevant for cold atomic gases, where the two-body effects have been shown to dominate the macroscopic response~\cite{Morice_1995, Andreoli_2021}. The obtained insight, however, can later be extended to cases where three or higher body phenomena could become relevant~\cite{Lehmberg_1970, Agarwal_2001,Ficek_2002}. 

\begin{figure}
    \centering
    \includegraphics[width=.9\linewidth]{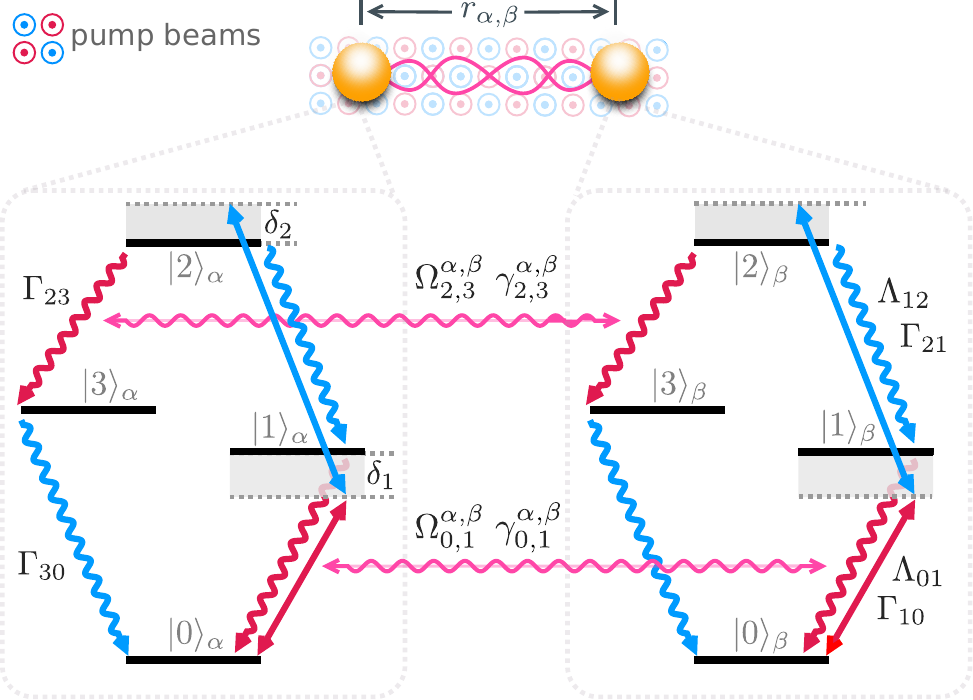}
    \caption{Energy-level diagram and schematics for two driven, four-level atoms ($\alpha$ and $\beta$) in the diamond configuration. Two pump beams detuned from the single- and two-photon transitions by $\delta_{1}$ and $\delta_{2}$ couple states $\vert0 \rangle$, $\vert 1 \rangle$, and $\vert 2 \rangle$ (straight arrows) with Rabi frequencies $\Lambda_{nm}$. De-excitation occurs via spontaneous emission (curled arrows) with individual decay rate $\Gamma_{nm}$. Blue and red lines denote two different polarizations. Atomic interactions are mediated by the electromagnetic environment, leading to a dipolar coupling and collective decay at rates $\Omega^{\alpha \beta}_{i,j}$ and  $\gamma^{\alpha \beta}_{i,j}$ (pink arrows). For simplicity only two of these channels are drawn. } \label{fig:four_level_atom}
\end{figure}

We consider two four-level atoms ($\alpha \in \{1,2\}$ and $m \in \{0,1,2,3 \}$), whose states are distributed in the diamond configuration, as shown in Fig.~\ref{fig:four_level_atom}. States $\vert 0 \rangle_{\alpha}$ and $\vert 2 \rangle_{\alpha}$ are connected through a two-photon transition induced by two external beams along state $\vert 1 \rangle_{\alpha}$ and a cascaded decay through $\vert 3 \rangle_{\alpha}$. The atoms interact with one another via environment mediated dipole-dipole transitions. The response is then determined by the competition of Rabi frequencies $\Lambda_{01}$ and $\Lambda_{12}$, detunings $\delta_{1}$ and $\delta_{2}$, and interatomic separation ${\mathbf{r}}_{\alpha \beta}$.

To be more specific, we have in mind four states of atomic  $^{87}$Rb: $\lbrace 5S_{1/2},\,5P_{3/2},\,5D_{3/2},\,5P_{1/2} \rbrace$,  which have been used to implement quantum control protocols and create narrow-band-squeezed light in alkali atomic gases \cite{Zibrov_2002, Becerra_2008, Willis_2010}. 

The states are separated by the transition wavelengths $\lbrace \lambda_{01},\, \lambda_{12},\, \lambda_{23}, \,\lambda_{30}\rbrace = \lbrace 780,\, 776, \,762, \,795 \rbrace\text{nm}$ and are connected by the dipole elements $\lbrace p_{0,1}, p_{1,2},p_{2,3},p_{0,3} \rbrace = \lbrace 5.956, 0.787,1.616,4.11 \rbrace \, ea_{0} $. This leads to the individual decay rates $\lbrace \Gamma_{0,1},\,\Gamma_{1,2},\,\Gamma_{2,3},\,\Gamma_{0,3} \rbrace = \lbrace 36.2,\,0.641,\,2.86,\,17.2\rbrace$\text{MHz}~\cite{Safronova_2004}. Since the transition energies differ significantly when compared to the decay rates, dipole-dipole effects are dominated by transitions between the same atomic levels, as illustrated by pink arrows in  Fig.~\ref{fig:four_level_atom}. At low temperatures ($\sim 100\mu$K) achievable in a  standard magneto-optical trap, the Doppler broadening for $^{87}$Rb atoms is about $0.1$~MHz, which is smaller than the decay widths of the involved transitions. 

The atoms are driven by two co-propagating pump lasers of variable powers and detunings. Their powers are chosen to give way to similar Rabi frequencies $(\Lambda_{01}\simeq \Lambda_{12})$, as done in recent experiments with atomic clouds~\cite{Cere_2018,Tellez_2022}. To better capture the competition between the driving fields and atomic separations, we report results for Rabi frequencies of the same order of magnitude of the highest individual decay rate $\Gamma_{01}$, with $\Lambda_{01}$ and $\Lambda_{12}$ ranging from $15$ to $45$\,MHz. With the dipole moments oriented as $\hat{p}_{0,1}=\hat{p}_{3,2}$ and $\hat{p}_{0,3}=\hat{p}_{1,2}$ while $\hat{p}_{0,1}^*\cdot\hat{p}_{1,2}=0$, a closed loop transition is guaranteed for $\hat{p}_{m,n}$ circularly polarized. The co-propagating configuration of the pump lasers further diminishes remaining Doppler effects by the selection of velocities mechanism.
Nevertheless, an ultracold sample (T$\sim 1\mu$K) similar to that achieved in~\cite{Bromley_2016} would avoid any significant Doppler effect without requiring a velocity selection scheme.

We consider it important to highlight two key differences between our description below and the standard approach used to depict the light generated by this system~\cite{Genes_2022}. First, we retain all atomic states throughout the analysis. The state $\vert 1 \rangle_{\alpha}$ is typically adiabatically eliminated due to the large detunings $\delta_{1}$ considered here~\cite{Brion_2007, Breuer_2007}. We demonstrate that, when dressed by the pump beams and dipole-dipole couplings, this state plays a significant role in the cascaded photons and, consequently, in the optical response of the system [Sec.~\ref{subsec:coll_dressed}]. Second, we explicitly state that there is no external beam connecting ~$\vert 3 \rangle_{\alpha}$ to the other states. It has been shown that the population inversion between levels $\vert 2 \rangle_{\alpha}$ and $\vert 3 \rangle_{\alpha}$ induced by the two-photon transition leads to lasing in a many-atom system~\cite{Malcuit_1985}. The amplified field can then be used to efficiently convert two low-power infrared laser fields into infrared and blue radiation in a rubidium gas through frequency up-conversion~\cite{Zibrov_2002}. The process is usually made efficient by closing the $\vert 3 \rangle_{\alpha}$ to $\vert 0 \rangle_{\alpha}$ transition via a weak seeding-field~\cite{Becerra_2008,Willis_2010}. We show here how atom-atom correlations give rise to a self-seeding field, leading to four-wave mixing where the cascade transition $\vert 2 \rangle_{\alpha} \rightarrow\vert 3 \rangle_{\alpha}\rightarrow\vert 0\rangle_{\alpha}$ exhibits coherence properties without an external pump field coupling states $\vert 3 \rangle_{\alpha}$ and $\vert 0\rangle_{\alpha}$.

\begin{figure}
    \centering
    \includegraphics[width=0.40\textwidth]{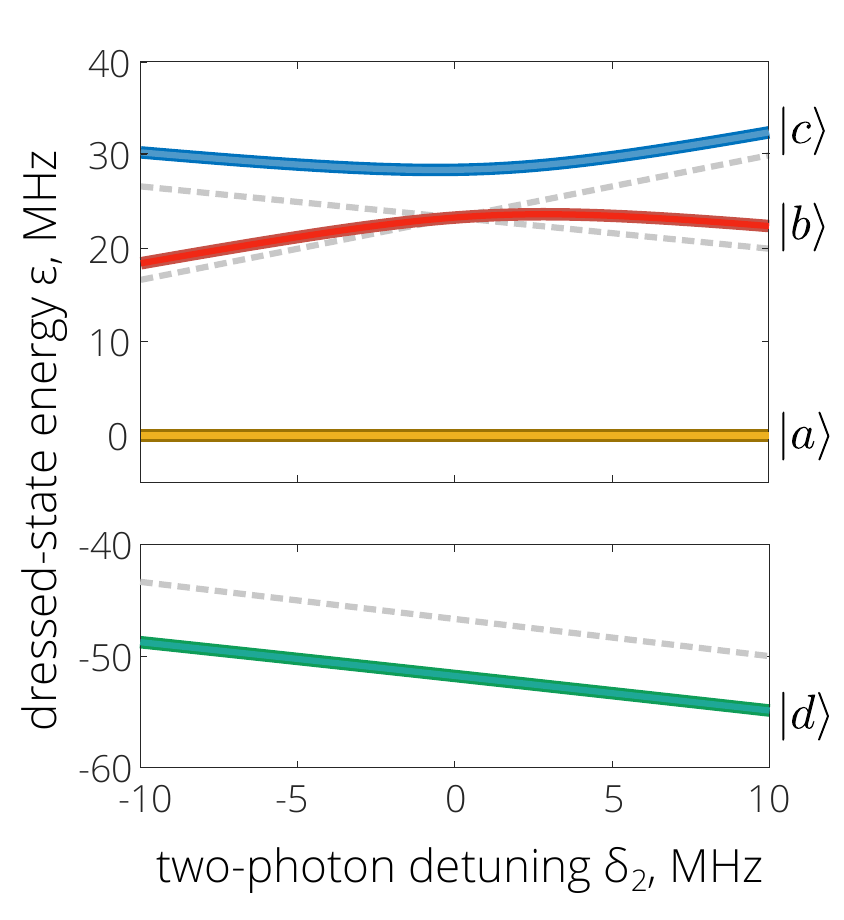}
    \caption{Dressed-state energies for a single atom in the diamond configuration as a function of the two-photon detuning $\delta_2$. The existence of several levels causes an additional shift between the bare energies (dashed lines) and dressed states at the crossing. For $\delta_1=-70 $~MHz and $\Lambda_{01},\Lambda_{12}=15$~MHz.}
    \label{fig:dressed-1atom}
\end{figure}

\subsection{Individual dressed states}\label{dressed2fourleveatoms}
We begin the FWM analysis  by reviewing the dressed states of an individual atom in the absence of decay. To do so, we move into an interaction picture with respect to the free Hamiltonian:
\begin{eqnarray}
\mathcal{H}_{0} &=& \mathcal{H}_{A} + \sum_\alpha\sum_{mn} \tfrac12 \hbar\varphi_{n m}\left(\sigma_{m,m}^\alpha
-\sigma_{n,n}^\alpha\right),
\label{eq:rotation}
\end{eqnarray}
where frequencies $\varphi_{01} = \frac{2}{3}(\delta_1+\delta_2)$, $\varphi_{12} = \frac{2}{3}(2\delta_2-\delta_1)$ are chosen to cancel out the laser frequencies. The resulting time-independent Hamiltonian within the bare basis is
\begin{equation}
\tilde{\mathcal{H}}_{\text{sys}} = \hbar\left(
\begin{array}{cccc}
-\frac{1}{3}\left(\delta_1+\delta_2\right) & -\Lambda_{0,1} & 0 & 0 \\
-\Lambda_{0,1} & \frac{1}{3}\left(2\delta_1-\delta_2\right) & -\Lambda_{1,2} & 0 \\
0 & -\Lambda_{1,2} & \frac{1}{3}\left(2\delta_2-\delta_1\right) & 0 \\
0 & 0 & 0 & 0 \\
\end{array}
\right).\label{eq:dressed1atom}
\end{equation}

Figure~\ref{fig:dressed-1atom} shows the dressed states that diagonalize $\tilde{\mathcal{H}}_{\text{sys}}$ as a function of the two-photon detuning $\delta_{2}$. The bare states $\vert 0 \rangle $ and $\vert 2 \rangle$ (dashed lines) are brought close to resonance via a two-photon transition. Due to the large single-photon detuning selected here, $\delta_{1} = -70$~MHz, the system transitions to the dressed states $\vert \mathrm{b} \rangle$ and $\vert \mathrm{c} \rangle$, which display an avoided crossing at two-photon resonance $(\delta_{2}=0)$ indicating a change in the role of the dominant bare state. By comparison, state $\vert 1 \rangle$ remains far from resonance and populates strongly the dressed state $\vert \mathrm{d} \rangle$, while $\vert 3 \rangle$  populates exclusively the dressed state  $\vert \mathrm{a} \rangle$. This last state remains decoupled from the others in the absence of spontaneous decay and we use it to define a zero point energy.  

In the case of two-photon resonance, individual dressed states acquire a simple form, providing insight into the collective effects as more atoms are added. When written within the bare basis, the individual dressed states read
\begin{subequations}\label{eq:dressed_1}
\begin{align}
\left\vert \mathrm{a}\right\rangle &= \left\vert 3\right\rangle, \\
\left\vert \mathrm{b}\right\rangle &\propto \Big[\Lambda_{1,2}\ket{0} - \Lambda_{0,1} \ket{2}\Big], \\
\left\vert \mathrm{c}\right\rangle &\propto \Big[\Lambda_{0,1}\ket{0} - \tfrac{1}{2} (\delta_1 + \zeta)\ket{1} + \Lambda_{1,2}\ket{2}\Big], \\
\left\vert \mathrm{d}\right\rangle &\propto  \Big[\Lambda_{0,1}\ket{0} - \tfrac{1}{2} (\delta_1 - \zeta)\ket{1} + \Lambda_{1,2}\ket{2}\Big],
\end{align}
\end{subequations}
where $\zeta=\sqrt{\delta _1^2+4 \Lambda _{0,1}^2+4 \Lambda_{1,2}^2}$. Their eigenvalues are 
\begin{subequations}\label{eq:dressed_2}
\begin{align}
& \varepsilon_\mathrm{a}=0, &
\varepsilon_\mathrm{b}=-\delta_1/3, \\
& \varepsilon_\mathrm{c}=(\delta_1+3\zeta)/6, &
\varepsilon_\mathrm{d}=(\delta_1-3\zeta)/6.
\end{align}
\end{subequations}
 From Eqs.~(\ref{eq:dressed_1}) it is shown that the state $\vert \mathrm{b}\rangle$ remains dark to an experiment probing the $\vert 1 \rangle$ state.

\begin{figure*}
\includegraphics[width=0.95\textwidth]{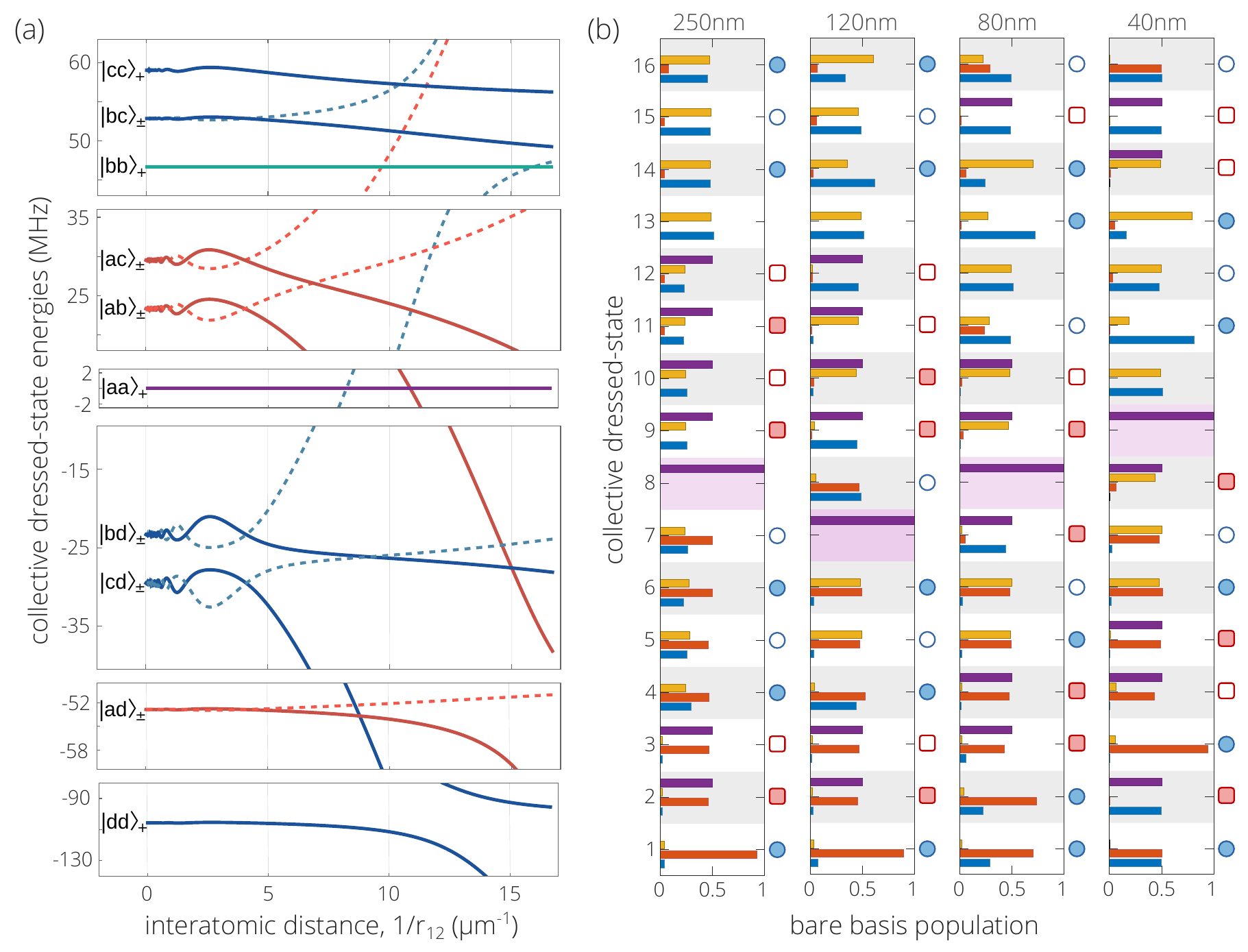}
\caption{Collective dressed states for two atoms located in a plane perpendicular to the laser propagation. (a) Dressed-state energies as a function of the atomic distance. The states organize into six uncoupled blocks denoted by blue ($\mathbb{BCD}_\pm$ block), red ($\mathbb{A}_{\pm}$), purple, and green lines. Solid and dashed lines denote symmetric and antisymmetric states respectively. Energy crossings occur among states of different blocks and avoided crossings among states of the same block. (b) Bare-state populations of the collective states obtained from Eq.~(\ref{eq:proj}), with blue, orange, yellow, and purple bars used for population inside $\vert 0 \rangle_{\alpha},\vert 1 \rangle_{\alpha}\vert 2 \rangle_{\alpha},\vert 3 \rangle_{\alpha}$. The dressed states are labeled from smaller to larger energies with blue circles and red squares used to denote states of the $\mathbb{BCD}_\pm$ and $\mathbb{A}_{\pm}$ blocks (empty shapes refer to antisymmetric states). The structure marks the change from the laser- to the dipole-dominated region as the atomic separation is reduced. Here  $\delta_1 = −70$~MHz, $\delta_2 = 0$~MHz, and $\Lambda_{01},\Lambda_{12}=15$\,MHz.}
    \label{fig:enter-label}
\end{figure*}

\subsection{Collective dressed states}\label{subsec:coll_dressed}
The individual dressed states allow a simple and accurate description of the atomic behavior when the atoms are far apart from one another and the dipole interactions $\mathcal{H}_{dd}$ can be neglected. 
{ In this regime, the atom pair is described by the symmetric and antisymmetric states of two identical isolated atoms
\begin{equation}\label{states_pm}
\ket{ij}_\pm =\frac{1}{\sqrt{2}}\Big[\ket{i}_1 \otimes \ket{j}_2 \pm \ket{j}_1 \otimes \ket{i}_2\Big] \, \, i,j = \mathrm{a, b, c, d}.
\end{equation}
The energies of each pair $\ket{ij}_\pm$ are then degenerate and are given by the sum of the individual dressed-state energies that, for $\delta_2=0$, are given by Eq.~(\ref{eq:dressed_2}).
As the atomic separation is reduced, these states  combine yielding  dressed states $\vert \upsilon\rangle$ as a superposition of $\ket{ij}_\pm$ states with coefficients that depend on the interatomic distance and the state-selective-coupling determined by driving fields and dipolar interaction. Nevertheless, the states Eq.~(\ref{states_pm})
remain a very convenient basis for the Hamiltonian. The dressed states  preserve their symmetry under atomic exchanges.
In the extreme case where the dipole-dipole interaction dominates over other interactions, including those involving the laser fields, the collective states can be naturally written in terms of the symmetrized superposition of two atom states built from $n,m \in\lbrace \mathrm{0, 1, 2, 3}\rbrace$ bare states. 

In this Section, we describe the evolution of the collective dressed states $\vert \upsilon \rangle$ and their corresponding eigenenergies when two-body effects are considered.}  

\subsubsection{Collective dressed-state energies}

 The underlying structure of the collective energy spectrum is significantly simplified by the selection rules of dipole transitions and invariance under atomic exchange. This becomes clearer when the Hamiltonian is expressed in terms of the $\ket{ij}_\pm$ states. In this basis, the selection rules separate the Hamiltonian into six independent blocks, thereby dividing the collective states into six decoupled regions. The first two blocks include the states $\vert \mathrm{aa} \rangle_{+}$ and $\vert \mathrm{bb} \rangle_{+}$, which remain decoupled for all atomic separations and $\delta_{2}=0$. The third block is formed by the collective states $\{\vert \mathrm{cc}\rangle_{+}, \vert \mathrm{dd}\rangle_{+}, \vert \mathrm{bc}\rangle_+,  \vert \mathrm{bd}\rangle_+,  \vert \mathrm{cd}\rangle_+\}$, which couple by the dipole terms. We coin this block as $\mathbb{BCD}_+$. The fourth block, $\mathbb{BCD}_{-}$, contains the states $\{\vert \mathrm{bc}\rangle_-,  \vert \mathrm{bd}\rangle_-,
 \vert \mathrm{cd}\rangle_-\}$. Finally, the fifth and six blocks,  $\mathbb{A}_\pm$, are formed by the symmetric and antisymmetric states  $\{ \vert \mathrm{ab}\rangle_\pm,  \vert \mathrm{ac}\rangle_\pm,\vert \mathrm{ad}_\pm\rangle\}$. 
 
 Once the division of the Hamiltonian into separate blocks is revealed, additional properties of the energy spectrum are understood. 
  This is illustrated in Fig.~\ref{fig:enter-label}(a)  where the energies of the collective dressed  states $\vert \upsilon\rangle$ and their corresponding bare-state population as a function of the interatomic distance are shown with parameters  consistent with experiments in atomic gases described previously. The two atoms lay on a plane perpendicular to the direction of propagation of the laser beams. The results are obtained by numerically diagonalizing the complete Hamiltonian at two-photon resonance ($\delta_{2}=0$) using the same parameters as in Fig.~\ref{fig:dressed-1atom}.

\subsubsection{Collective dressed-state populations}

The energy spectrum underscores the importance of the far-detuned state $\vert 1 \rangle_{\alpha}$. This bare state mediates dipole transitions between $\vert 0 \rangle_{\alpha}$ and $\vert 2 \rangle_{\alpha}$ states. Had it been adiabatically eliminated, transitions to the asymptotic states $\vert \mathrm{bc}\rangle_{+}$ and $\vert \mathrm{cc} \rangle_{+}$ would be dipole forbidden. The small changes in their lines as the atomic separation is varied shows the importance of keeping this state and the coherence induced by the laser even at small distances.  We now examine how the bare state populations of the collective states as a last step before describing how they are connected by the decay channels. 

Figure~\ref{fig:enter-label}(b) shows the bare-state populations of the collective states $\vert \upsilon \rangle$ with $\upsilon =1,...,16$ for different atomic separations. The states are arranged by increasing energy ($\varepsilon_\mathrm{\upsilon+1}>\varepsilon_\mathrm{\upsilon}$) and their populations are obtained via
\begin{equation}
 \mathcal{P}^{\upsilon}_m= \text{Tr}\left[\vert m\rangle_{1} \leftindex_{1}\langle m\vert \upsilon\rangle \langle  \upsilon\vert \right] {{= \text{Tr}_{2} [\vert \langle \upsilon \vert m \rangle_{1} \vert^{2} ]}}\, ,
\label{eq:proj} 
\end{equation}
where $\vert m \rangle_{\alpha=1}$ are the bare states { which have a straightforward interpretation}.The same results are obtained for $\alpha =2$ due to the invariance under atomic exchange. 

The populations are represented as colored bars, with the color scheme consistent with Fig. \ref{fig:dressed-1atom}. The uncoupled state $\vert \mathrm{\mathrm{aa}} \rangle = \vert 3 3 \rangle$ highlighted in purple as a guide to the eye. At large separations $r_{12}=250$~nm, the populations correspond to those of the individual dressed states [see Eq.~(\ref{eq:dressed_1})]. As the separation is reduced to  $\kappa_{mn}r_{1,2} \approx 1$, the dipole-dipole interaction starts to strongly couple individual dressed states. It becomes of the same order of magnitude as the Rabi interaction around this separation and dominates as the separation is further reduced. The coupling depends on the involved wavelength, as heralded by the states $\vert \upsilon=9 \rangle$ and $\vert \upsilon=12 \rangle$ at $r_{12}=120$~nm, which become mostly populated by $\vert m=0,3 \rangle_{1}$, the two bare states that are separated by the largest wavelength $\lambda_{03}=795$~nm. The two collective states approach the symmetric and antisymmetric pairs of the dipole coupled bare states as described above for short distances. Additional pairs begin to form as the distance is further reduced, as shown at $40$~nm where states $\vert \upsilon = 1 \rangle$ and $\vert \upsilon = 16 \rangle$ represent symmetric and antisymmetric pairs, as well as $\vert \upsilon = 2 \rangle$ and $\vert \upsilon = 15 \rangle$. Close to an avoided-crossing, the corresponding populations of a pair of ``energy-colliding" states are similar and show a mixture between individual dressed and bare states. For example at $80$~nm where both pairs $\vert \upsilon = 1 \rangle$-$\vert \upsilon = 2 \rangle$ and  $\vert \upsilon = 10 \rangle$-$\vert \upsilon = 16 \rangle$  are close to an avoided crossing their states present similar populations, a population that is not easily described in terms of bare or individual dressed states.

We have marked the states of the blocks $\mathbb{BCD}_{\pm}$ by blue circles and those of $\mathbb{A}_{\pm}$ by red squares in Fig.~\ref{fig:enter-label}(b). Empty shapes denote antisymmetric states while filled ones denote symmetric ones to guide their relation as distance is reduced. In addition, as mentioned above, states belonging to $\mathbb{BCD}_{\pm}$ maintain the coherence induced by the laser through the $\vert 1 \rangle_{\alpha}$ state at short distances and their distinction will prove to be important when describing the correlations between the photons generated in the cascaded decay via $\vert 3 \rangle_{\alpha}$ in the next section.

Given a value of the Rabi frequencies $\Lambda_{mn}$ , calculations have been performed for the energy and composition of the collective dressed states with different orientation of the relative separation of the atoms. The qualitative behavior is similar. The main quantitative difference corresponds to the
values of $r_{12}$ at which the dipole-dipole interaction is first non-negligible and latter become dominant.  This feature is a direct consequence of the anisotropic behavior of $\Omega_{n,m}^{\alpha\beta}$. For the case under study, which involves dipole-dipole terms associated with circularly polarized transitions, 
the magnitude of $\Omega_{n,m}^{\alpha\beta}$ exhibits a minimum
at $\sin \theta \sim \sqrt{2/3}$, and the conservative dipole-dipole interaction becomes non-negligible at the shortest interatomic separations.

\subsection{Collective decay rates}

The collective dressed states are ultimately connected via decay channels obtained by diagonalization of the decay tensor $\gamma^{\alpha \beta}$ of Eq.~(\ref{eq:similar_energies}). These channels determine the rate at which photons are scattered by the atoms and the path an excitation follows as it leaves the ensemble.

\begin{figure}[hbt!]
    \centering
    \includegraphics[width=0.48\textwidth]{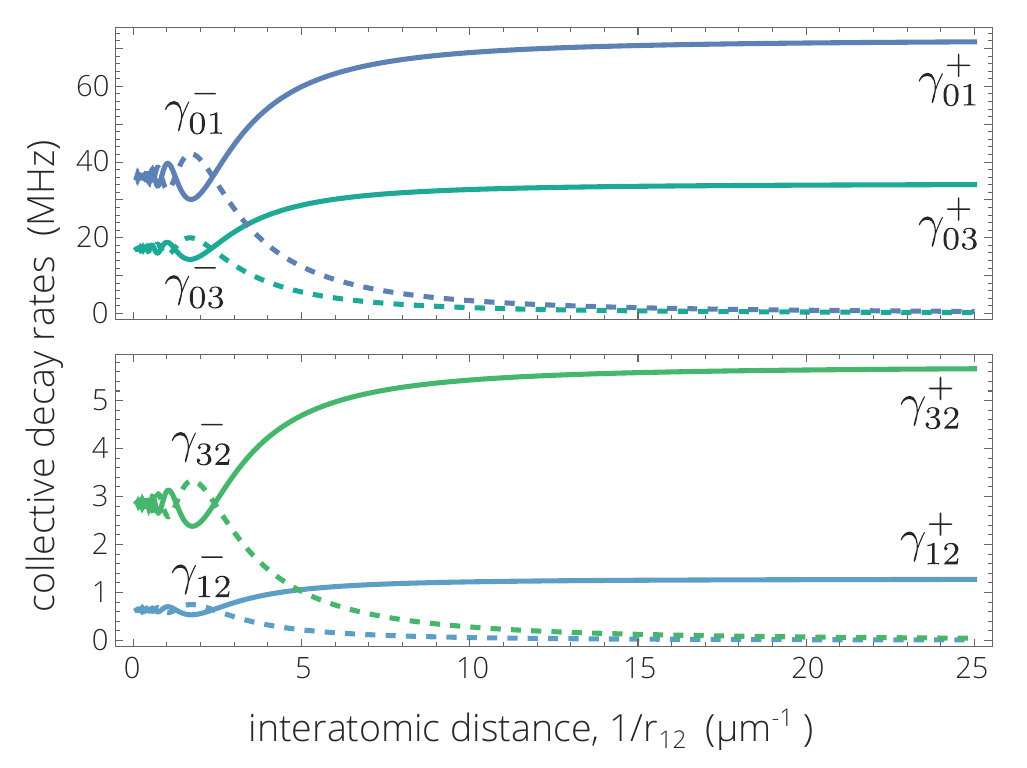}
 \caption{Collective decay rates $\gamma^{\pm}_{m,n}$ as a function of the interatomic distance for two atoms located in a plane perpendicular to the laser propagation. At large distances, super~$(+)$ and sub-radiant~$(-)$ contributions converge to the individual decay rate. At short distances the super-radiant contribution saturates to twice the individual decay rate and the sub-radiant contribution vanishes. The general behavior is similar as the orientation of $\mathbf{r}_{1,2}$ is varied, causing only slight deviations in the amplitude of the oscillations around $r_{12}>200$~nm.} \label{fig:decayrates}
\end{figure}

There are two decay channels for each dipole-allowed $(m,n)$-transition: a symmetric, super-radiant channel with decay rate $\gamma_{mn}^+=\gamma_{mn}^{1,1}+ \gamma_{mn}^{1,2}$; and an antisymmetric, sub-radiant channel, with decay rate $\gamma_{mn}^-=\gamma_{mn}^{1,1}- \gamma_{mn}^{1,2}$. Figure~\ref{fig:decayrates} shows the decay rates as the atomic separation is changed for two atoms in the same plane perpendicular $\theta = \pi/2$. The decay rates display oscillations around the individual decay rates $\gamma_{mn}$ for large inter-atomic distances $\kappa_{m,n}r_{12}>1$, before saturating to twice the individual decay rate for super-radiant cases and vanishing to zero for sub-radiant cases. These oscillations are similar to those encountered in the energies plotted in Fig.~\ref{fig:enter-label} and discussed in Appendix~\ref{ap:Bessel_functions}.  Results are qualitatively and quantitatively similar for other relative orientations of the atoms due to the smooth dependence of $\gamma_{n,m}^{\alpha,\beta}$ on  $r_{1,2}$ and on $\sin^2\theta=(\hat {\mathbf{p}}^*\cdot\hat {\mathbf{r}})
(\hat {\mathbf{p}}\cdot\hat {\mathbf{r}})$.

The effect of the collective decay channels over the atomic ensemble is accounted for by the jump operators
\begin{equation}\label{eq:pi+-}
\Pi_{mn}^\pm =\frac{1}{\sqrt{2}}\big(\sigma_{mn}^1 \otimes \mathbbm{1}_{2} \pm \mathbbm{1}_{1}\otimes\sigma_{mn}^2 \big) \, . 
\end{equation}
The operators are organized according to their influence on the blocks introduced in the previous section. The jump operators $\Pi_{01}^\pm$ and $\Pi_{12}^\pm$ go through the intermediate state $\vert 1 \rangle_{\alpha}$. Super-radiant jumps couples states within the same block, thus allowing for the emitted photons to be traced back to atomic transitions among the excitation ($\mathbb{BCD}_{\pm}$ to $\mathbb{BCD}_{\pm}$) or cascaded ($\mathbb{A}_{\pm}$ to $\mathbb{A}_{\pm}$) blocks. Sub-radiant jumps, by comparison, mix states of different symmetry, but keep them inside the excitation or cascaded blocks. That is, they transfer states from $\mathbb{BCD}_{\pm}$ to $\mathbb{BCD}_{\mp}$ or  $\mathbb{A}_{+}$ to $\mathbb{A}_{-}$. This symmetry changing transitions become less common as the atomic separation is reduced. The remaining jump operators $\Pi_{03}^\pm$ and $\Pi_{32}^\pm$ go through the intermediate state $\vert 3 \rangle_{\alpha}$ and complement this response. They mix states of the excitation and cascaded blocks. Super-radiant jumps mix blocks of the same symmetry ($\mathbb{BCD}_{\pm}$ and $\mathbb{A}_{\pm}$), while sub-radiant jumps mix blocks of opposite symmetry ($\mathbb{A}_{\pm}$ and $\mathbb{B}_{\mp}$).

\section{Collective effects imprinted on individual and photon pairs}\label{subsec:g2}

We have shown that the collective dressed states of two atoms reflect the competition between Rabi frequencies $\Lambda$ and dipole-dipole couplings $\Omega$, leading to three regions of interest: (i) laser dominated; (ii) dipole dominated; and (iii) intermediate.  We have also introduced the collective decay channels and presented systematically how they mix the different states. In this final Section we discuss how the correlations of the photons scattered by the atoms change as we transition between the three regions. Different regions can be accessed by changing the laser powers or the atomic separation.  The following calculations bring our analysis closer to experiments where correlations between photon pairs are controlled externally by the pump beams in a scenario where two-body interactions dominate, as, for example, in cold gases~\cite{Morice_1995, Andreoli_2021}.

To characterize the outgoing photons we consider the first- and second-order correlation functions:
\begin{eqnarray}
   \mathcal{G}^{(1)}(x_1) &=& \expval{E^\dagger(x_1) E(x_1)} \, ,
    \label{eq:G1}\\
    \mathcal{G}^{(2)}(x_1,x_2) &=& \expval{E^\dagger(x_1) E^\dagger(x_2) E(x_2)E(x_1)} \, ,
    \label{eq:G2}
\end{eqnarray}
written here in terms of the electric field amplitudes. The functions are proportional to the probability of detecting single photons or photon pairs at the spacial-temporal coordinates $x_{i}=(\boldsymbol{R}_{i},t_{i})$. Photons can arrive from the lasers or the atoms as presented in Eqs.~(\ref{eq:Free_fields})-(\ref{eq:positive}), where the field was divided into the free and scattered terms $ \mathbf{E}^{(+)} = \mathbf{E}_{\text{f}}^{(+)} +  \mathbf{E}_{\text{s}}^{(+)}$. The scattered terms relate to the atomic variables via
\begin{multline}
\mathbf{E}_{\text{s}}^{(+)}(\boldsymbol{R},t_{i})=
\frac{\hbar}{\epsilon_0c^2}
\sum_\alpha\sum_{\lbrace n,m \rbrace}
\frac{\boldsymbol{p}_{mn}
-\hat{\boldsymbol{R}}_\alpha(\hat{\boldsymbol{R}}_\alpha
\cdot \boldsymbol{p}_{mn})}{R_\alpha}\\
\times
\Delta_{mn}^2 \exp\left[i\kappa_{mn}\left(R_\alpha-R\right)\right]
\sigma_{mn}^{\alpha}(t) \, ,
\label{eq:farfield5}
\end{multline}
in the far field region, where $t_R = t+R/c$ is thee retarded time and $\boldsymbol{R}_\alpha = \boldsymbol{R}-\boldsymbol{r}_\alpha$ the distance between the probing point $\boldsymbol{R}$ and the atom $\alpha$. As above, the sum $\lbrace n,m \rbrace$ is restricted to transitions from upper to lower levels only.

We are interested on photons emitted along the cascaded decay $\vert 2\rangle\rightarrow \vert 3 \rangle\rightarrow \vert 0\rangle$ and thus filter out the laser photons and those scattered from the excitation branch. To collect the scattered photons, we consider photon detectors placed at $\mathbf{R}=(R,\theta,\phi)$ in the far field region. Each detector covers a surface area ${\mathcal A} = R^{2} \Delta\Omega$ with solid angle $\Delta\Omega$. For simplicity, we assume that photons are emitted in the direction of the laser propagation ($\hat R \cdot \vec p_{23} = \hat R\cdot \vec p_{30} = 0$). Photons are very likely to be scattered in this direction due to phase-matching conditions. 

The probability of detecting a photon emitted from the $(m\rightarrow n)$-transition is $\mathcal{G}^{(1)}_{m,n}{\mathcal A}
    :=  \mathcal{K}_{mn} \text{G}^{(1)}_{m,n}$ with 
\begin{multline}
\text{G}^{(1)}_{mn} =
    \expval{\sigma^{1}_{mm}} + \expval{\sigma^{1}_{mn}\sigma^{2}_{nm}} \mathrm{e}^{i\kappa_{mn}\hat{\boldsymbol{R}}
     \cdot \boldsymbol{r}_{1,2}}\\
   +\expval{\sigma^{2}_{mm}}
   +\expval{\sigma^{1}_{nm}\sigma^{2}_{mn}} \mathrm{e}^{-i\kappa_{mn}\hat{\boldsymbol{R}}
     \cdot \boldsymbol{r}_{1,2}}\, ,
       \label{eq:G1-ij}
    \end{multline}
and $\mathcal{K}_{mn}=[(\hbar^2 \omega^4_{mn})/(\epsilon_0 c^2)] \vert {\mathbf p}_{mn}\vert^2 \Delta \Omega$. The probability to detect two-photon coincidences radiated in the cascaded decay within a small time window is $\mathcal{G}^{(2)}_{2,3;3,0}{\mathcal A}^2:=  \mathcal{K}_{23}\mathcal{K}_{30} \text{G}^{(2)}_{2,3;3,0}$, with 
\begin{multline}
\text{G}^{(2)}_{2,3;3,0}= 
    \expval{\sigma^{1}_{22}}
   +\expval{\sigma^{2}_{22}}
   +\expval{\sigma^{1}_{22}\sigma^{2}_{33}}
   +\expval{\sigma^{1}_{33}\sigma^{2}_{22}}\\
   +2\expval{\sigma^{1}_{32}\sigma^{2}_{23}}
   \mathrm{e}^{i\kappa_{32}\hat{\boldsymbol{R}}
     \cdot \boldsymbol{r}_{1,2}}
   +2\expval{\sigma^{1}_{23}\sigma^{2}_{32}}
   \mathrm{e}^{-i\kappa_{32}\hat{\boldsymbol{R}}
     \cdot \boldsymbol{r}_{1,2}}\\
   +\expval{\sigma^{1}_{02}\sigma^{2}_{20}}
      \mathrm{e}^{i\kappa_{02}\hat{\boldsymbol{R}}
     \cdot \boldsymbol{r}_{1,2}}
   +\expval{\sigma^{1}_{20}\sigma^{2}_{02}}
   \mathrm{e}^{-i\kappa_{02}\hat{\boldsymbol{R}}
     \cdot \boldsymbol{r}_{1,2}},
     \label{eq:G2(0)}
\end{multline}
and $\kappa_{02}=\kappa_{32}+\kappa_{03}$. 

Below we calculate the functions $\text{G}^{(1)}_{mn}$ and $\text{G}^{(2)}_{2,3;3,0}$ for different regions of the parameter space. The functions depend on the expectation values of atomic operator strings in the interaction picture, obtained by numerically solving Eq.~\eqref{eq:Qdot13}, with the atoms initially prepared in the ground state. The evolution is tracked until a steady-state is reached at $t\sim800$\, ns for the assumed parameters. { Details of the computation that involve the expectation values of the adequate operators are provided in Appendix~\ref{ap:sollehmberg}. An alternative to perform such calculations make use of a reduced density matrix \cite{James_1993,Keaveney_2012,Javanainen_2014,Jenkins_2016,Zwanzig_1961,Kiffner_2007,Zhou_2009}.}

\subsection{Two-atom effects on single photon detection}

\begin{figure}[hpt!]
\centering
        \includegraphics[width=0.475\textwidth]{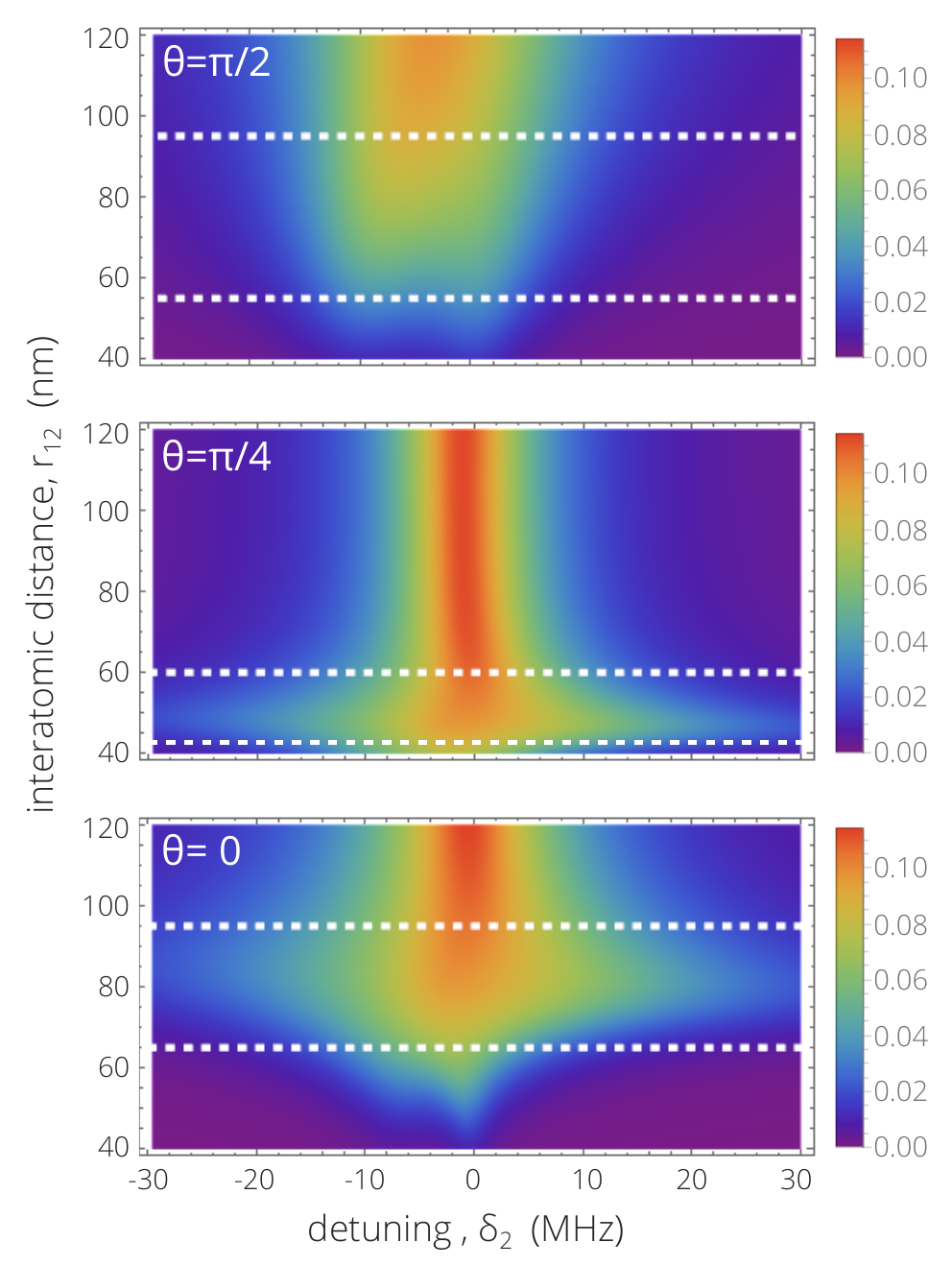}
    \caption{First order correlation $\text{G}^{(1)}_{30}$ of photons radiated by the $\vert 3\rangle-\vert 0\rangle $ transition, as a function of detuning $\delta_2$ and atomic separation $r_{12}$. The plots are shown for three different angles formed between the vector connecting the atoms $\mathbf{r}_{12}$ and the laser propagation $\mathbf{k}_{q}$. The maximum value is obtained at large distances and is similar across all angles. As the separation is lowered the function changes appreciably as explained in the main text. These changes mark the transition between laser-dominated, intermediate, and dipole-dominated regions, whose boundaries are plotted as white dotted lines. The regions are determined by the dressed-state energies, they depend on the ratio between Rabi frequency and dipole coupling and, as such, change for each angle. Here $\delta_1 = −70$\,MHz and  $\Lambda_{01}=\Lambda_{12} = 15$\,MHz. }
    \label{fig:G130}
\end{figure}

Figure~\ref{fig:G130} shows the first-order correlation $\text{G}^{(1)}_{30}$ as a function of atomic separation and laser detuning. The figure is divided into three panels distinguished by $\theta$, the angle between the line connecting the atoms $\mathbf{r}_{\alpha\beta}$ and the laser propagation direction $\boldsymbol{k}_{q}$. In all cases, the correlation function displays a significant change as we transition between the laser- and dipole-dominated regions. Where and how these changes occur varies for different angles, as the dipole interaction $\Omega$ varies with $\theta$ and nearly vanishes for $\theta = \pi/4$  [see Eqs.~(\ref{eq:tensorF}) and~(\ref{eq:Green_Function})]. 
In the figure, the dashed lines mark
the boundaries between the different regions. The positions of the lines are obtained from the avoided crossings in the energy spectrum, as plotted for $\theta=\pi/2$ in Fig.~\ref{fig:enter-label}. The first avoided crossing marks the end of the laser-dominated region while the last one marks the beginning of the dipole-dominated one. The positions of the lines change for each angle. Such boundaries are selected only as a guide, as the changes between regions occur smoothly. However, they help to highlight qualitative changes in the correlation function.

At large distances the correlation function displays a single peaked distribution for all angles. The distribution resembles that obtained for the population of the state $\vert 2\rangle$ of an isolated three-level atom in the ladder configuration interacting with the two-pump fields.
In fact, for the weak driving fields considered here, the analytical expressions of Ref.~\cite{Whitley_1976,Cere_2018} predict a maximum value of $\langle\sigma_{22}^{\alpha}\rangle$ at $\delta_2 \sim -2$~MHz with a power-broadened spectrum of full-width at half maximum (FWHM) around $11$~MHz, in accordance with out results. The model also predicts that observing an Autler-Townes doublet, as a result of the laser beams~\cite{Willis_2007}, would require much greater power than the values considered here.

One reason both the diamond and ladder configurations lead to similar correlation functions $\text{G}^{(1)}_{23}$ and $\text{G}^{(1)}_{30}$ at large distances is that their behavior is primarily dominated by the population in the upper state $\vert 2 \rangle_{\alpha}$. Since the population in the intermediate state $\vert 3 \rangle_{\alpha}$ largely mimics the behavior of the population of $\vert 2 \rangle_{\alpha}$ , the structure of both  $\text{G}^{(1)}_{23}$ and $\text{G}^{(1)}_{30}$ differ essentially by a scale factor that changes slightly (about 10$\%$) for different orientations.

As the atomic distance is reduced, the joint coherences $\expval{\sigma^{1}_{23}\sigma^{2}_{32}}$ and $\expval{\sigma^{1}_{30}\sigma^{2}_{03}}$ become relevant and coherence-effects induced on the photons from the atomic interactions appear. These effects are evident in Fig.~\ref{fig:G130} at short distances ($r_{1,2} \lesssim70$~nm), where the correlations functions for $\theta=0, \pi/2$ show a double peaked structure that vanishes outside $-6<\delta_2<2$\,MHz. The effect appears at shorter distances for $\theta = \pi/4$ as coherent dipole interactions are reduced at this atomic orientation. The distance at which these effects become relevant is further reduced when the power of the lasers is increased.

\begin{figure}[hpt!]
\centering
        \includegraphics[width=0.485\textwidth]{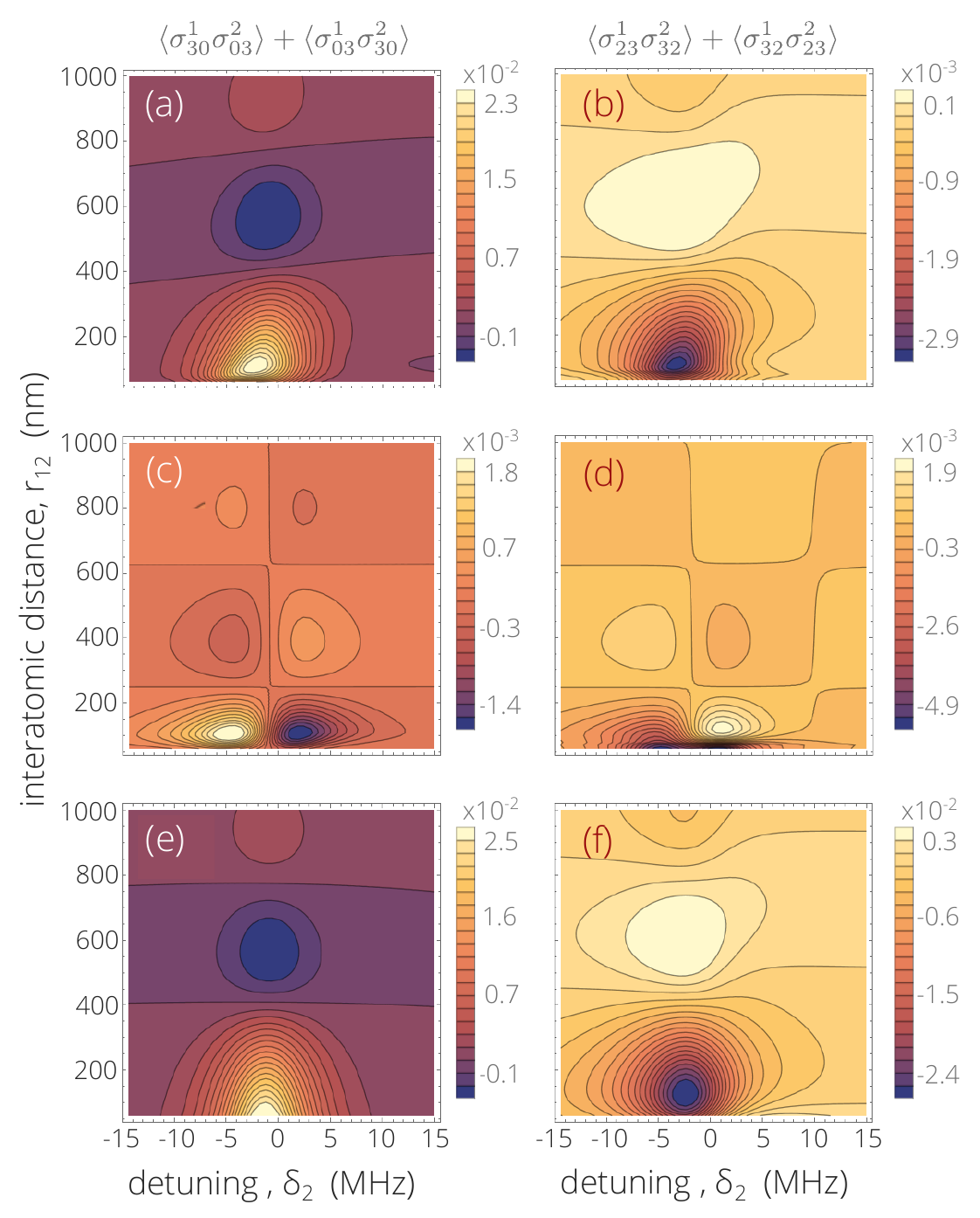}
\caption{Expectation values for the inter-atomic coherences  $\expval{\sigma^{1}_{23}\sigma^{2}_{32}}_+$ (first column) and $\expval{\sigma^{1}_{03}\sigma^{2}_{30}}_+$ (second column) for two atoms located in a plane perpendicular to the propagation direction of the pump lasers, as a function of the detuning $\delta_2$ and the inter-atomic separation $r_{1,2}$.  (a-b) All couplings are included. (c-d) Collective decay rates are $\gamma^{1,2}_{m,n}$ artificially turned off. (e-f) Dipole couplings $\Omega^{1,2}_{m,n}$ are artificially turned off. For $\delta_1 = −70$\,MHz,  $\Lambda_{01},\Lambda_{12}=15$\,MHz. }
    \label{fig:08}
\end{figure}

There are two standard mechanisms for coherence to arise in the diamond configuration: population inversion and self-seeding. For photons radiated from the transition $\vert 2 \rangle$-$\vert 3 \rangle$ it can be caused by a population inversion between these two states. For photons radiated from transition $\vert 3 \rangle$-$\vert 0 \rangle$ this inversion is not possible, and the coherence can be attributed to a self-seeding effect. We plot the expected values of $\expval{\sigma^{1}_{23}\sigma^{2}_{32}}$ and $\expval{\sigma^{1}_{03}\sigma^{2}_{30}}$ coherences in Figs.~\ref{fig:08}(a)-(b) for $\theta=\pi/2$, where the atomic plane is perpendicular to the laser propagation. Both coherences reach their maximum absolute value near $\delta_2=-2$~MHz and $r_{1,2}\sim 100$~nm, close to the intermediate region. The coherences exhibit opposite signs, which accounts for the initial difference between $\text{G}_{23}^{(1)}$ and $\text{G}_{30}^{(1)}$ at short distances. The second change arises from the steady state population $\langle \sigma^{\alpha}_{33} \rangle$ and coherence $\langle \sigma^{1}_{30} \sigma^{2}_{03} \rangle$, which are of the same order of magnitude in this regime, highlighting the significance of dipole-dipole correlations. Figure~\ref{fig:08}(c)-(f) are used to distinguish whether the coherences arise from the dipole-dipole interaction or the collective decay channels. For panels~(c) and~(d) the latter are turned off while leaving individual decays. For panels~{{(e) and~(f)}} the coherent dipole-dipole interactions are turned off. The similarity between the full solution and the one without coherent couplings indicates that the coherences predominantly arise from collective decays.

\subsection{Two-atom effects on cascade photon detection}

\begin{figure*}[hpt!]
\centering
    \begin{tabular}{ccc}
        \includegraphics[width=0.95\textwidth]{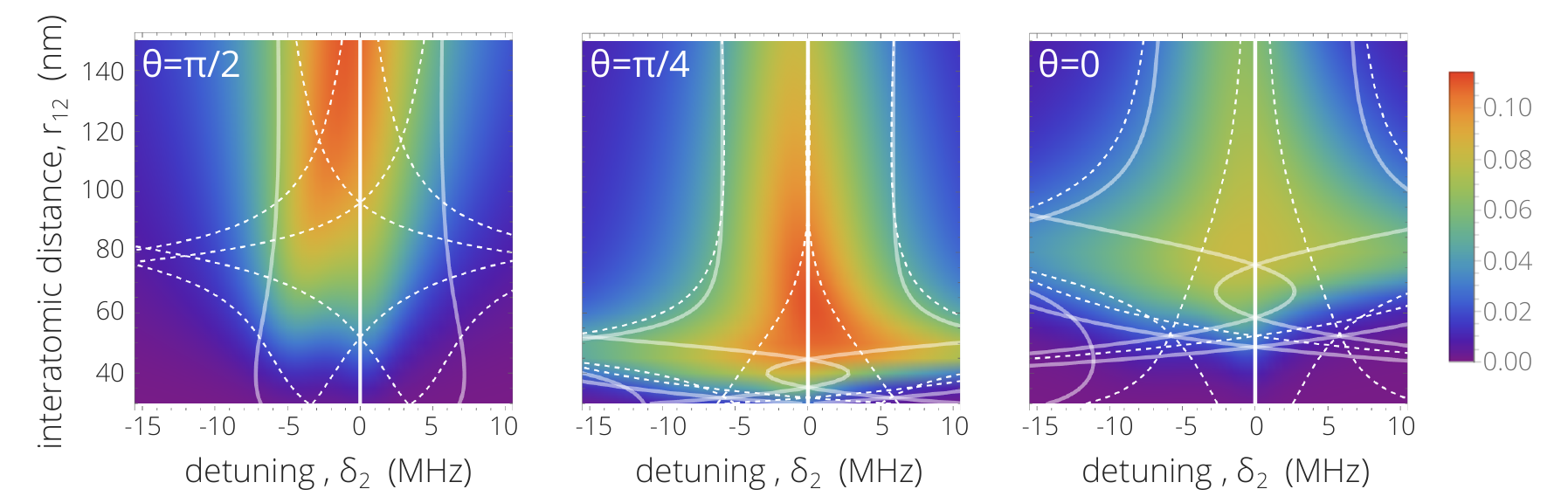}
    \end{tabular}
    \caption{Second order correlation $\text{G}^{(2)}$ of photons radiated along the cascade transition $\vert2 \rangle-\vert 3 \rangle-\vert 0 \rangle$ as a function of the detuning $\delta_2$ and the atomic separation $r_{1,2}$. The overall behavior is accurately described by transitions among three relevant collective dressed states as explained in the text. Solid white lines mark the energy splitting between two dressed states connected by a superradiant transition, while dashed lines mark the splitting between those connected by subradiant transitions. Across the laser-dominated and intermediate regions the behavior is ruled by superradinat transitions. Inside the dipole-dominated region, however, the system becomes trapped around the ground state for the weak Rabi frequencies considered here $\Lambda_{01},\Lambda_{12} = 15$\,MHz and subradiant channels dominate. The angles $\theta$ and detuning $\delta_{1}$ are the same as in Fig.~\ref{fig:G130}.}
    \label{fig:G2_supsub}
\end{figure*}

The driving lasers create coherent paths between the states $\vert 0 \rangle_{\alpha}$ and $\vert 2 \rangle_{\alpha}$ to give way to the individual dressed states of Eq.~(\ref{eq:dressed_1}). These coherent paths are also imprinted in the cascade branch through the collective decays and dipole-dipole interactions, and they are readily represented in the second-order correlation function through terms of the form $\expval{\sigma^{1}_{02}\sigma^{2}_{20}}$, $\expval{\sigma^{1}_{20}\sigma^{2}_{02}}$ in Eq.~(\ref{eq:G2(0)}). We now show how the competition between the laser field, dipole interaction, and collective decay is reflected in measurements of the
correlation function $\text{G}^{(2)}_{2,3;3,0}$.
In doing so, we illustrate the predictive power of the analysis presented above.

Figure~\ref{fig:G2_supsub} shows the second order correlation $\text{G}^{(2)}_{2,3;3,0}$ as function of atomic separation and laser detuning for three different orientations of the vector connecting the atoms. Superimposed on each plot are white lines obtained from the collective dressed states and decay channels, as explained below. The overall shape of the correlation function appears to resemble that of $\text{G}^{(1)}$ plotted in Fig.~\ref{fig:G130}. It shows a single-peaked distribution that degrades as the atomic separation is reduced, eventually giving way to a double peaked distribution readily observable for $\theta=0,\pi/2$. The transition from single to double peaked distribution is caused by dipole-dipole interactions.  In the absence of dipole-dipole interactions, states $\vert n\rangle_1\vert m\rangle_2$ and $\vert m\rangle_1\vert n\rangle_2$ are degenerate. These states are coupled by the dipole interaction $\Omega_{nm}^{\alpha\beta}$, creating a pair of states whose energies become separated  
and yield the double peaks. The physical mechanism is similar to that occurring in cavities when a vacuum mode and an
atom are considered as two otherwise degenerate oscillators  whose coupling leads to hybrid
modes with specific energies, a phenomenon known as vacuum Rabi splitting. 
This approach has been famously used to explain the Mollow spectrum in terms of the dressed states of a driven two-level atom, where two symmetric peaks appear at the standard Rabi splitting~\cite{Cohen_1989}. In fact, our  results are compatible with the observation of collective multimode vacuum Rabi splitting in free space already reported in~\cite{Guerin_2019}.

Having this in mind, the energy spectra of the dressed states and the collective decay channels are now used to understand the  dependence of the cascaded photons coincident emission on the two-photon detuning and the atomic separation. We have evaluated the energy differences among all the dressed states in our model and found that the behavior of three levels appear to dominate the ${\text G}^{2}(0)$ spectrum. These levels have the largest contribution of the states~$\vert 0 \rangle_{\alpha}$ and $\vert 2 \rangle_{\alpha}$, allowing for transitions between the ground and upper state. They belong to the $\mathbb{BCD}_\pm$ blocks discussed in Sec.~\ref{sec:twofourlevatom} and are found in the frequency interval $\delta_{2} \in [40,60]$ in Fig.~\ref{fig:enter-label}. The state $\vert bb \rangle$ also yield similar energy differences. However, it predominantly exhibits the character of a dark state because the two strongest decay channels ($\Gamma^+_{0,1}$ and $\Gamma^+_{0,3}$) do not lead to transitions to $\ket{\mathrm{bb}}$, and the subradiant channels are too weak to produce noticeable decays in this parameter space (see Fig. \ref{subsec:g2}).

The energy differences $(\varepsilon_\mathrm{\upsilon}-\varepsilon_\mathrm{\upsilon'})$ between the three relevant states are plotted as white lines Fig.~\ref{fig:G2_supsub}. The six lines represent all possible energy differences among these levels. Solid lines are used to represent super-radiant transitions, where the two relevant states share the same symmetry, while dashed lines represent sub-radiant transitions, where the symmetry changes (see Sec.~\ref{sec:twofourlevatom}~C).
It can be observed that the plotted lines delimit the correlation function for all the possible orientations of the atomic vectors. This occurs regardless of the strong differences between the $\text{G}^{(2)}$ functions for different angles. The lines also provide insight into the system dynamics as we transition between laser- and dipole-dominated regions. Inside the intermediate and laser-dominated regions, the most relevant transitions are super-radiant ones, causing an enhancement of photons emitted in transition between these states. Once at the dipole-dominated region, the super-radiant decay rate becomes much larger than the Rabi frequencies (for our system $\gamma_{01}^{+}\simeq 72$~MHz). Such a strong decay rate inhibits the excitation of the system causing a large population of the ground state $\vert 00\rangle$. For $\theta =\pi/4/$ and $\theta =0$
the super-radiant transitions yield an increase of ${\text{G}}^{(2)}(0)$ with respect to its isolated atom value at the intermediate region. At smaller inter-atomic distances and for all orientations, the structure of ${\text{G}}^{(2)}(0)$ in this regime is found to be then determined from the sub-radiant transitions. 

Weak laser fields are required to reveal the dynamics within the dipole-dominated region at the distances accessible in current experimental settings. We have performed similar calculations for other values of the Rabi frequencies
$\Lambda_{01}=\Lambda_{12}\leq \Gamma^+_{0,1}$, which yield similar results. As the Rabi frequencies increase, Rabi broadening becomes more pronounced, and cooperative effects play a more prominent role at shorter distances. If $\Lambda_{01}=\Lambda_{12}> \Gamma^+_{0,1}$, even at very short distances, the populations of the excited states $\vert 1\rangle$ and $\vert 2\rangle$ will not be negligible, and the two-peak structure is lost.

\section{Conclusions}\label{sec:Conclusions}

The dressed atom formalism has been a staple of quantum optics studies that provides a natural picture to explain the quantum behavior of atoms coupled to classical or quantum fields. In this work we have considered the collective dressed states of an ensemble of multi-level atoms, where the atoms are dressed by both external photons and those exchanged with neighboring atoms. These collective dressed states where then used to explain the optical response of two four-level atoms in the diamond configuration, a problem with strong implications for nonlinear optics in alkali atomic gases. By working inside the dressed-state picture we were able to unveil three interaction regions: (i) laser-dominated; (ii) dipole-dominated; and (iii) intermediate, and characterize the response at each region.

The connection between the atomic state and the optical response was explored in these regions via the first and second order correlation functions of the outgoing photons. We determined the structure of these correlations in terms of the atomic separation and laser detuning and found that, for nearby atoms, the response can be controlled by the external lasers only. Using the example of $^{87}$Rb, we found that for weak lasers (Rabi frequency $\Lambda_{01}=\Lambda_{12}=15$\,MHz comparable to individual decay rates $\Gamma_{01}\simeq 36.2 $~MHz) the spectrum can change from single to double peaked, to reflect the transition between laser- and dipole-dominated regions for atomic separations achievable in current experimental settings. By working inside the dressed-state formalism we delineated how the excitation branch (atomic levels $\vert 0 \rangle-\vert 1\rangle-\vert 2\rangle$) could alter photons emitted in the cascaded decay ($\vert 2 \rangle-\vert 3\rangle-\vert 0\rangle$). These results highlight the coherent and collective effects that arise between several atoms and can not be obtained through individual ones. We also highlighted the importance of the intermediate level $\vert 1 \rangle$ which is adiabatically eliminated in standard descriptions of this system.

Macroscopic effects found in ordered atoms or dense atomic gases can be approached from the collective response already found for the two atoms considered here. In particular, the energy splitting found here, is expected found for atomic samples where a significant number of atomic pairs are found at separations $r_{12}< 80$\, nm for the parameters used here. These results are compatible with both the observation of collective multimode  vacuum Rabi splitting reported in \cite{Guerin_2019} and the structured $\text{G}_2(0)$ coincidence functions observed and reported in \cite{Cere_2018}. These experiments involve atomic gases with general parameters compatible with our calculations albeit with atoms that do not have a constant inter-atomic separation.

Our results also act as a proof of principle that self-seeding is a cooperative effect that can be induced by a single pair of atoms at an adequate distance. More detailed calculations involving statistical distributions of inter-atomic distances are needed to quantitatively test whether
self-seeding appears in an atomic cloud. The feasibility of measuring the macroscopic  manifestation of this microscopic effect is supported by the stability of the relevant energy differences between the associated dressed states as functions of the inter-atomic distance for $\kappa_{nm}r_{12}< 1$.

The connection with this different results, illustrates how our calculations can be used in scenarios where the  dipole-dipole interactions dependent on the distance and orientation between atoms, competes with the action of external lasers to create inter-atomic correlations with clear consequences for the nonlinear optical observables.

\section*{ACKNOWLEDGEMENTS}
RG-J, DS and RJ gratefully acknowledge support by the PAPIIT-UNAM grants IA103024, IN106821, IN112624, and IN104523, { and CONAHCYT CF-2023-G15}. AK is indebted to IFUNAM for their hospitality and was financially supported by Departamento de Ciencias B\'asicas UAM-A grant number 2232218.  PY-T was financially supported by CONAHCYT Estancias Posdoctorales por M\'exico 2022 (3). PB-B gratefully aknowledges support from the PAPIIT-UNAM grant IG101324. We thank LSCSC-LANMAC where part of the computational simulations involved in this work were performed in their hpc server.
\appendix

\section{Closer look at the dressed-states energies} \label{ap:Bessel_functions}

\begin{figure*}[hbt!]
    \centering
    \includegraphics[width=0.32\textwidth]{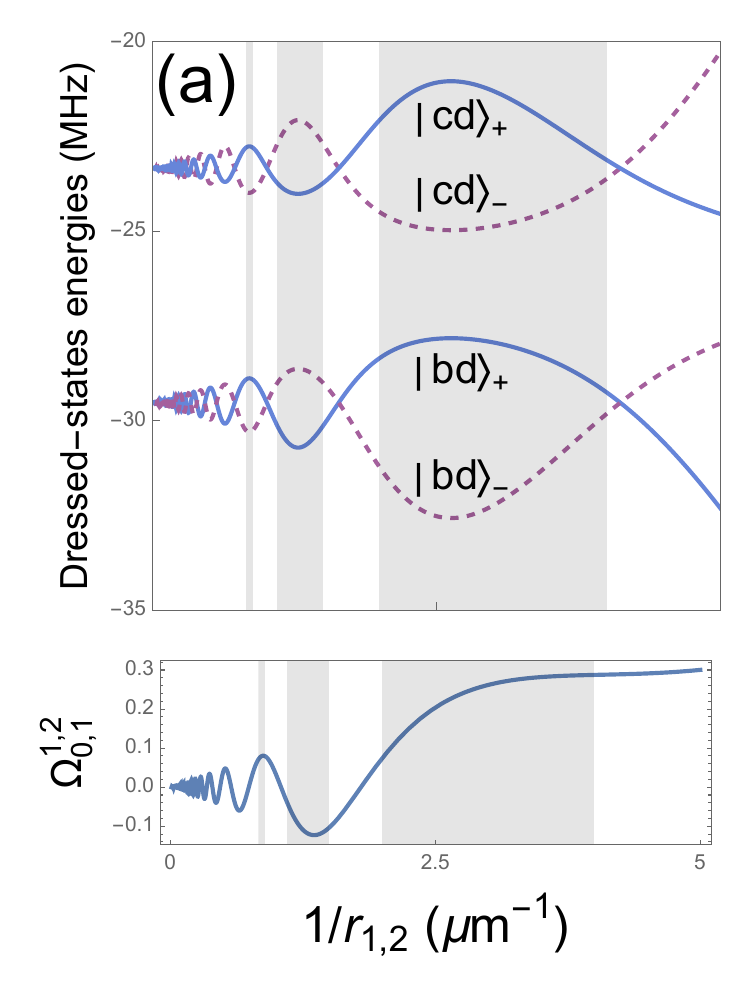}
    \includegraphics[width=0.32\textwidth]{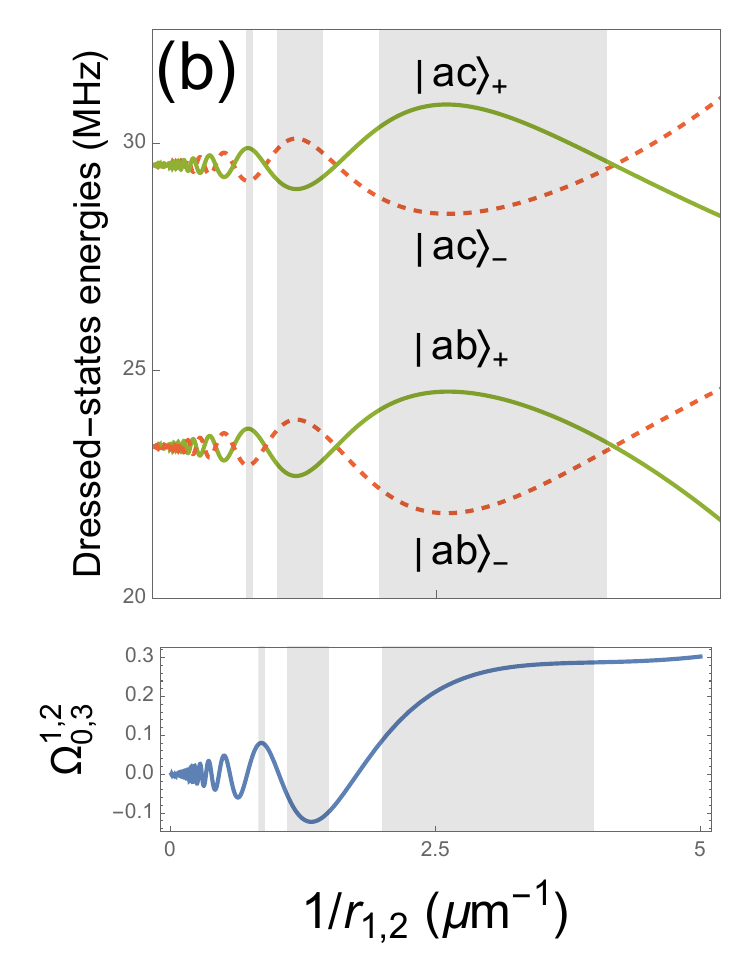}
    \includegraphics[width=0.32\textwidth]{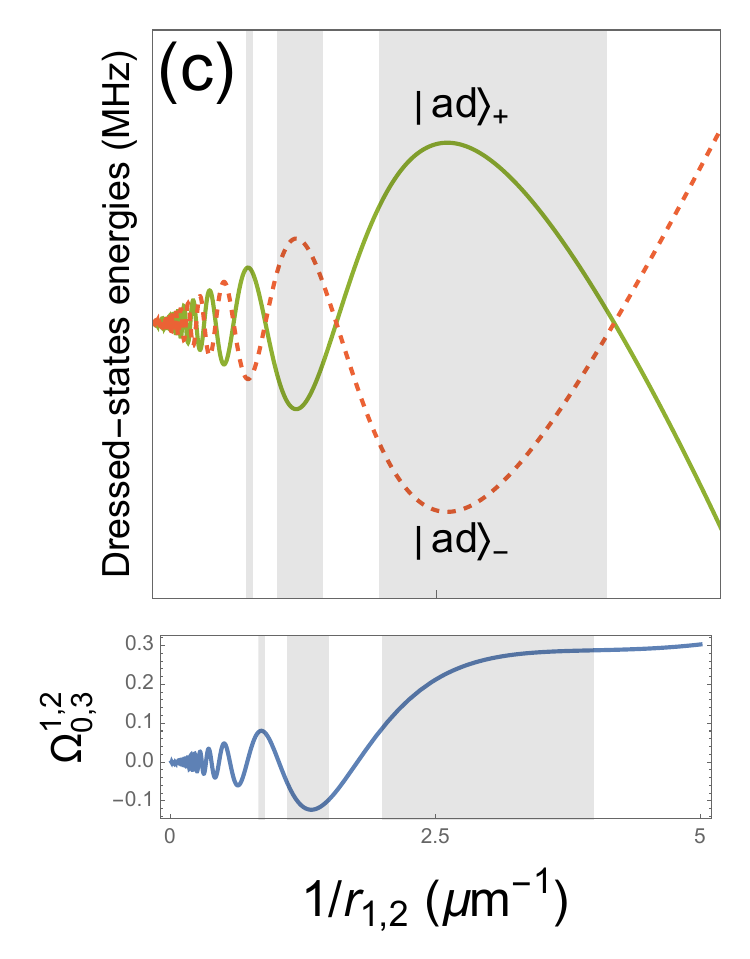}
    \caption{Energies of the collective dressed states of two atoms as a function
     of the inverse of the inter-atomic distance for $\delta_1=-70$~MHz and
     $\delta_2=0$~MHz, $\Lambda_{01}=15$\,MHz and $\Lambda_{12}=15$\, MHz.
     Traces of the oscillatory behavior of $\Omega^{1,2}_{mn}$
     inherited from the spherical Bessel functions composing the Green function
     $\overleftrightarrow{\mathbf{G}}(\mathbf{r},\omega)$are observed. 
     In this particular case all involved wavelengths are very similar, around 780~nm.}
    \label{fig:dressed-2atom-b}
\end{figure*}

In the particular example that we work out in detail, the Rabi frequencies are similar, $\Lambda_{0,1}\sim \Lambda_{1,2}$, $\vert\delta_1\vert\gg\Lambda_{1,2}$ and the electric dipoles are such that 
$\vert \vec{p}_{01}\vert\gg \vert \vec{p}_{12}\vert$ and $\vert \vec{p}_{30}\vert \gg \vert \vec{p}_{23}\vert $.
At large distances ($\kappa_{mn} r_{1,2}\gg 1$) the interaction consists mainly of photon exchanges associated with the $1/\kappa_{mn} r_{1,2}$ term which dominates the behavior of $\Omega^{12}_{m,n}$ in the radiation zone [see Eq.~(\ref{eq:Green_Function})]. At short inter-atomic distances this tensor acquires a much richer structure that strongly modifies the dressed state energies. The effect is greater as the distance is reduced and the quasi static
term $1/\kappa_{mn}^2 r_{1,2}^3$ dominates. Finally, the dipole approximation begins to break down as $r_{1,2}$ approaches the atomic size~\cite{Anderson_2013}.

The oscillatory changes of the two-atom energy are illustrated in 
Figure \ref{fig:dressed-2atom-b} and compared to the behavior of the predominant couplings $\Omega_{mn}^{1,2}$ for different states as a function of the inter-atomic distance. In this numerical simulation the atoms are assumed to be located in the same plane perpendicular to the main direction of propagation of the laser beams. Similar results are obtained for other atomic configurations. In Figure \ref{fig:dressed-2atom-b}(a), $\Omega_{0,1}^{1,2}$ is kept on and all other coupling are artificially turned off. We observe the lifting of the degeneracy for the states 
$\ket{\mathrm{bd}}_\pm$ and $\ket{\mathrm{cd}}_\pm$. Similarly, in Figs. \ref{fig:dressed-2atom-b}(b) and \ref{fig:dressed-2atom-b}(c)  only $\Omega_{0,3}^{1,2}$ is kept on and the degeneracy is lifted only from the states labeled $\ket{\mathrm{ab}}_\pm$, $\ket{\mathrm{ac}}_\pm$ and $\ket{\mathrm{ad}}_\pm$, these collective states all involve the single atom dressed state $\ket{\mathrm{a}}$ which corresponds asymptotically to the bare state $\ket{3}$. Each $\Omega^{1,2}_{m,n}$ coefficient is responsible for lifting some degeneracies, and only in the case when all coupling are kept on are all the degeneracies lifted. The modified energies follow the same shape as the dominant $\Omega_{mn}^{1,2}$.
Symmetric and antisymmetric states are affected
so that, for $r_{12} > 350$ nm
$(1/r_{12} <2.85 \mu\text{m}^{-1}$),
the states that asymptotically correspond to 
$\vert \mathrm{c}i\rangle_\pm$ and $\vert \mathrm{b}i\rangle_\pm$, 
($i = \{\mathrm{a,d}\}$), show essentially the same oscillations in their energy shifts but with
opposite sign. This similarity is attributed to the structure of $\vert \mathrm{b}\rangle$ and $\vert \mathrm{c}\rangle$ states of Eqs.~\eqref{eq:dressed_1}.

\section{Solution of the dynamical equations} \label{ap:sollehmberg}

In this Section we describe a general approach to solve Eq. \eqref{eq:Qdot13}, written in short-hand notation as
\begin{equation}
\dot{Q}(t)=L\left(Q(t),t\right) \, . \label{eq:dynQ}
\end{equation}
Any atomic operator $Q(t)$ can be written as a linear combination
of the elements of the set of matrices $\mathcal{Q}_N=\{Q_1, Q_2,\dots, Q_N\}$
that spans the whole set of atomic operators.
The operator $\dot{Q}(t)$ as well as the operator
$L\left(Q(t),t\right)$ in Eq.~(\ref{eq:dynQ}) can also be
written in terms of such a basis.
Since the total number of states for a system of emitters
is $n_t=n_l^{n_a}$, the set $\mathcal{Q}_N$ must
be formed of $N=n_t^2=n_l^{2n_a}$ matrices.

It can be shown that any such group of matrices
forms an internal vector space and can be
arranged so that they are orthogonal under the trace
scalar product
\begin{equation}
\mathrm{Tr}\left[Q_i^\dagger Q_j\right]=
\mathrm{Tr}\left[Q_i^\dagger Q_i\right]\delta_{i,j}.
\label{eq:orthogonal0}
\end{equation}
Thus, any atomic operator $A$
can be expanded in terms of  the
elements of $\mathcal{Q}_N$ as
\begin{equation}
A=\sum_i c_iQ_i,
\label{eq:Aexpand0}
\end{equation}
where $c_i=\mathrm{Tr}[Q_i^\dagger A]/\mathrm{Tr}[Q_i^\dagger Q_i]$,
according to the orthogonality
condition \eqref{eq:orthogonal0}.

There is a wide variety of sets
that can be used as the basis $\mathcal{Q}_N$.
To construct a
complete
set of matrices of dimension $n_t$
we first build a
complete set of matrices of
dimension $n_l$ that span the space
of a single emitter. Altogether this set
will have $n_l^2$ elements.
The elements of this set belong to one of three categories
of matrices. 
The first $n_l$ matrices
$q_1$, $q_2$, $\dots$, $q_{n_l}$
are  diagonal matrices
with only one diagonal element equal to $1$ and all other elements equal to
$0$. These are given explicitly by
\begin{equation}
\left(q_k\right)_{mn} = \delta_{m,k}\delta_{n,k},
\end{equation}
where $i,m,n = 1,2,\dots, n_l $.
Next is a set of matrices that
are generalizations of the $\sigma_x$
Pauli matrix. These are give by
\begin{eqnarray}
(q_k)_{mn} &=& \delta_{m,j}\delta_{n,l}+\delta_{m,l}\delta_{n,j},
\end{eqnarray}
where $j=1,2,\dots, n_{l}-1$, $l=m+1,m+2,\dots, n_l$
and $k=n_l+1, n_l+2,\dots, n_l(n_l+1)/2$.
The final matrices in the third set are generalizations of the
$\sigma_y$ Pauli matrix. Similarly to the
previous set, these are give by
\begin{eqnarray}
(q_k)_{mn} &=& i\left(\delta_{m,j}\delta_{n,l}
-\delta_{m,l}\delta_{n,j}\right),
\end{eqnarray}
where $j=1,2,\dots, n_{l}-1$, $l=m+1,m+2,\dots, n_l$
and $k=n_l(n_l+1)/2+1, n_l(n_l+1)/2+2, \dots, n_l^2$.
We can now build the complete base as
a Kronecker product of the $q_k$ matrices
\begin{equation}
Q_k=q_{k_{n_a}}\otimes q_{k_{n_a-1}}\otimes \dots \otimes q_{k_1},
\end{equation}
where $k_1, k_2, \dots, k_{n_a}=1,2,\dots, n_l^2$
and $k=\sum_{i=0}^{n_a-1}k_in_l^{i}$.
This base has the advantage of being
composed of Hermitian matrices therefore
any Hermitian matrix expanded in this base
will have exclusively real coefficients.

We can write $N$ differential
equations, one for each one of the elements of
$\mathcal{Q}_N$ as
\begin{equation}
\frac{d}{dt}Q_i(t)
=L\left(Q_i(t),t\right).
\label{eq:MLEexpand0}
\end{equation}
The formal solution to this equation is
\begin{equation}
Q_i(t) = U(t, Q_i),
\label{eq:nonunitevol}
\end{equation}
where $U$ is a linear map.

This allows us to expand the right-hand side
of \eqref{eq:MLEexpand0} as
\begin{equation}
\frac{d}{dt}Q_i(t)
=\sum_{j}\frac{\mathrm{Tr}\left[Q_j^\dagger L\left(Q_i,t\right)\right]}
{\mathrm{Tr}\left[Q_j^\dagger Q_j\right]}Q_j(t).
\label{eq:MLEexpand1}
\end{equation}

Now lets assume that initially the ensemble of emitters
is in the state
\begin{equation}
\left\vert \Psi_0\right\rangle=
\left\vert \psi_{n_a}\right\rangle\otimes \left\vert \psi_{n_a-1}\right\rangle
\otimes \dots \otimes\left\vert \psi_1\right\rangle,
\end{equation}
where $\left\vert\psi_i\right\rangle$
is the initial state of the $i-$th emitter.
Taking the expectation value of both sides
of \eqref{eq:MLEexpand1}
and defining $w_i(t) = \left\langle \Psi_0 \left\vert Q_i(t)
\right\vert \Psi_0 \right\rangle$
we obtain
\begin{equation}
\frac{d}{dt}w_i(t)
=\sum_j\mathrm{Tr}\left[Q_j^\dagger L\left(Q_i,t\right)\right]
w_j(t).
\label{eq:MLEexpand2}
\end{equation}
Defining the vector of expectation values
$\boldsymbol{w}(t)=(w_1(t), w_2(t), \dots, w_N(t))$
we note that this procedure yields quite a
simple linear system of differential equations
of the form
\begin{eqnarray}
\frac{d \boldsymbol{w}(t)}{dt} = \Lambda(t)\boldsymbol{w}(t),
\end{eqnarray}
with initial conditions
$w_i(0)= \left\langle \Psi_0 \left\vert Q_i
\right\vert \Psi_0 \right\rangle$
and  matrix elements given by $\Lambda_{i,j}(t)
=\mathrm{Tr}\left[Q_j^\dagger L\left(Q_i,t\right)\right]$.
Furthermore, the matrix elements of $\Lambda$ are
all real numbers, because, as we mentioned above,
$\mathcal{Q}_N$ is constituted of Hermitian matrices.

Once the $w_i(t)$ are known,
the expected value of any
operator $A$ can be expressed
by using expansion \eqref{eq:Aexpand0} as
\begin{equation}
\left\langle \Psi_0 \left\vert A(t)
\right\vert \Psi_0 \right\rangle 
= \sum_i
\frac{\mathrm{Tr}\left[Q_i^\dagger A\right]}
{\mathrm{Tr}\left[Q_i^\dagger Q_i\right]}w_i(t).
\end{equation} In this manner one need only solve Eq.~\eqref{eq:Qdot13} once for a given set of parameters and can employ that solution to calculate the expectation value of any arbitrary atomic operator.

\end{document}